\documentclass[fleqn,usenatbib]{mnras}

\usepackage{newtxtext,newtxmath}

\usepackage[T1]{fontenc}

\DeclareRobustCommand{\VAN}[3]{#2}
\let\VANthebibliography\thebibliography
\def\thebibliography{\DeclareRobustCommand{\VAN}[3]{##3}\VANthebibliography}

\usepackage{graphicx}	
\usepackage{subcaption}
\usepackage{amsmath}	
\usepackage{tabularx}
\usepackage{hyperref}
\usepackage{rotating}
\usepackage{threeparttable}
\usepackage{mathrsfs} 
\usepackage{xcolor}   
\usepackage{array}
\usepackage{multirow} 
\newcommand{\MBH}{M_\mathrm{BH}}
\newcommand{\MBHm}{\langle M_\mathrm{BH}^{3\sigma} \rangle}
\newcommand{\Mbh}{M_\mathrm{BH}}
\newcommand{\Mstar}{M_\mathrm{\star}}
\newcommand{\Mh}{M_\mathrm{h}}
\newcommand{\fduty}{f_{\mathrm{duty}}}
\newcommand{\Msun}{M_{\odot}}
\newcommand{\NH}{N_\mathrm{H}}
\newcommand{\Mbulge}{M_\mathrm{bulge}}

\newcommand{\BHAR}{\langle {{\mathrm{BHAR}}^{3\sigma}} \rangle}
\newcommand{\SFR}{\langle \mathrm{SFR} \rangle}
\newcommand{\HAR}{\langle \mathrm{HAR} \rangle}
\newcommand{\Mhm}{\langle M_\mathrm{h} \rangle}
\newcommand{\LXXRB}{L_{\mathrm{X,XRB}}}
\newcommand{\LXAGN}{\langle{L_{\mathrm{X,AGN}}^{3\sigma}}\rangle}

\newcommand{\Lbol}{L_\mathrm{bol}}
\newcommand{\LbolAGN}{L_\mathrm{bol,AGN}}
\newcommand{\LbolAGNm}{\langle L_\mathrm{bol,AGN}^{3\sigma} \rangle}
\newcommand{\lamedd}{\lambda_\mathrm{edd}}
\newcommand{\MUV}{M_\mathrm{UV}}
\newcommand{\Lxm}{{\langle L_{2-10 \mathrm{keV}}^{3\sigma}}\rangle}
\newcommand{\z}{\langle z \rangle}

\title[X-ray stacking reveals SMBH accretion of SFGs]{X-ray stacking reveals average SMBH accretion properties of star-forming galaxies and their cosmic evolution over $4 \lesssim z \lesssim 7$}

\author[Matsui et al.]{
Suin Matsui$^{1}$\thanks{E-mail: \href{smatsui@astron.s.u-tokyo.ac.jp}{smatsui@astron.s.u-tokyo.ac.jp} 
},
Kazuhiro Shimasaku$^{1,2}$,
Kei Ito$^{1}$,
Makoto Ando$^{1}$,
Takumi S. Tanaka$^{1,3,4}$
\\
$^{1}$Department of Astronomy, Graduate School of Science, The University of Tokyo, 7-3-1 Hongo, Bunkyo-ku, Tokyo 113-0033, Japan\\
$^{2}$Research Center for the Early Universe, Graduate School of Science, The University of Tokyo, 7-3-1 Hongo, Bunkyo-ku, Tokyo 113-0033, Japan\\
$^{3}$Kavli Institute for the Physics and Mathematics of the Universe, The University of Tokyo, Kashiwa, Chiba 277-8583, Japan\\
$^{4}$Center for Data-Driven Discovery, Kavli IPMU (WPI), UTIAS, The University of Tokyo, Kashiwa, Chiba 277-8583, Japan
}

\date{Accepted XXX. Received YYY; in original form ZZZ}

\pubyear{2023}

\begin{document}
\label{firstpage}
\pagerange{\pageref{firstpage}--\pageref{lastpage}}
\maketitle

\begin{abstract}
%
With an X-ray stacking analysis of $\simeq 12,000$ Lyman-break galaxies (LBGs) 
using the Chandra Legacy Survey image,
we investigate average 
supermassive black hole (SMBH)
accretion properties of star-forming galaxies 
(SFGs)
at $4\lesssim z\lesssim7$. Although no X-ray signal is detected in any stacked image, we obtain strong $3\sigma$ upper limits for the average black hole 
accretion rate (BHAR) as a function of star formation rate (SFR).
At $z \sim 4$ 
(5) where the stacked image is deeper, the $3\sigma$ BHAR upper limits per SFR are $\sim 1.5$ ($1.0$) dex lower than the local black hole-to-stellar mass ratio, indicating that the SMBHs of SFGs in the inactive (BHAR $\lesssim 1 M_\odot$ yr$^{-1}$) phase are growing much more slowly than expected from simultaneous evolution. We obtain a similar result for BHAR per dark halo accretion rate.
QSOs from the literature
are found to have $\sim 1$ dex higher SFRs and $\gtrsim 2$ dex higher BHARs than LBGs with the same dark halo mass.
We also make a similar comparison for dusty starburst galaxies and quiescent galaxies 
from the literature.
A duty-cycle corrected analysis shows that for a given dark halo, the SMBH mass increase in the QSO phase dominates over that in the much longer inactive phase. 
Finally, a comparison with the TNG300, TNG100, SIMBA100, and EAGLE100 simulations finds that they overshoot our BHAR upper limits by $\lesssim 1.5$ dex, possibly implying that simulated SMBHs are too massive.
\end{abstract}




\section{Introduction}
\label{sec:intro}
A tight correlation between the supermassive black hole (SMBH) mass ($M_\mathrm{BH}$) and the 
bulge mass ($\Mbulge$)
seen in local galaxies suggests that SMBHs and galaxies have co-evolved \citep{Kormendy2013, McConnell2013}.
When did the correlation emerge?
What physical mechanisms made this correlation? 
To answer these questions and unveil the history of co-evolution, it is needed to investigate the correlation between the two over cosmic time. 

Beyond the local universe, however, observations are limited to active galactic nuclei (AGNs) because it is challenging to measure $M_\mathrm{BH}$ for inactive galaxies 
owing to the absence of broad lines.
It is also hard to measure $\Mbulge$ for distant AGNs because of the difficulty in imaging their morphology, and thus, 
the total stellar mass ($\Mstar$) is often used as a proxy of it.
Previous studies of $z>0$ AGNs (including QSOs) have obtained mixed results that may depend on redshift. 
Studies up to $z \sim 2.5$ have found no significant evolution of the $\MBH$--$\Mstar$ relation \citep[e.g.,][]{Ding2020, Suh2020, Li2021} {\footnote{\citet{Ding2020} have also measured $\Mbulge$ and detected positive evolution of $\MBH/\Mbulge$.}}.
At higher redshift, SMBHs tend to be overmassive with respect to the local relation \citep[e.g.,][]{Bogdan2023, Harikane2023, Maiolino2023,Ubler2023}, 
although there have been found faint sources whose $M_\mathrm{BH}/\Mstar$ is consistent with the local ratio \citep[e.g.,][]{Ding2023, Kocevski2023, Larson2023}.

An important problem in studies using AGNs 
is that the obtained results may be biased because AGNs are rare objects with high accretion rates.
We cannot rule out the possibility that AGNs, in particular QSOs, are outliers of the correlation scattered toward high $M_\mathrm{BH}$ \citep[e.g.,][]{Lauer2007}.
Future observations of fainter sources will provide a less biased view of the correlation.

Given the difficulties 
in measuring 
$\MBH$ at $z>0$, an alternative, although indirect, method for studying co-evolution that can also be applied to 
inactive galaxies 
is to use the growth rates of SMBHs and stellar components, i.e., BH accretion rate (BHAR) and star formation rate (SFR). 
If the local $M_\mathrm{BH} - \Mstar$ relation holds over cosmic time, BHAR/SFR ratios will be close to an average $M_\mathrm{BH}/\Mstar$ value.
Indeed, a similarity in the evolution of 
the star formation rate density (SFRD) and the BH accretion rate density (BHAD) seen at $z \lesssim 2.5$ \citep[e.g.,][]{Delvecchio2014, Vito2018} 
suggests
some connection between star formation and BH accretion.
Note, however, that this similarity does not necessarily mean the simultaneous evolution of galaxies and SMBHs on an individual basis because the SFRD and BHAD are integrated quantities over all galaxies.

BHARs at $z>0$ can be estimated from X-ray luminosities with less contamination from star formation than at other wavelengths. Even for galaxies whose X-ray emission is too faint to be detected, we can obtain an average BHAR by stacking their X-ray images and discuss co-evolution in an unbiased way.
Stacking analysis has an additional advantage that it essentially provides a time-averaged BHAR of a given galaxy sample by smoothing out possible variability of individual sources on short timescales of $10^2$ -- $10^7$ yr \citep[e.g.,][]{Novak2011, Hickox2014}
{\footnote{Short time-scale variability is an uncertain factor when one uses the AGN X-ray luminosity function to constrain cosmological simulations that do not have sufficient time resolution \citep{Habouzit2022}.}}.

The BHAR -- SFR relation is of particular interest at high redshifts where most galaxies, including massive ones, are star-forming galaxies (SFGs) 
\citep[e.g.,][their Fig.16]{Davidzon2017}.
In this study, we examine this relation 
for 
SFGs at $4 \lesssim z \lesssim 7$ by X-ray stacking.

Several papers have examined the relation between BHAR and host galaxy properties including SFR at 
redshifts similar to our redshift range 
by X-ray stacking of a large sample.
\citet{Fornasini2018} have used $\simeq 75,000$ SFGs at $0.1<z<5$ in the Chandra COSMOS-Legacy Survey area to find that while BHAR strongly correlates with $\Mstar$, there is no correlation with SFR. \citet{Carraro2020} have conducted a similar analysis on three types of galaxies (SFGs, starburst galaxies [SBGs], and quiescent galaxies [QGs]) at $0.1 < z < 3.5$ of a total number of 97,746, 
finding that BHAR/SFR depends on 
$\Mstar$
and that it does not evolve with redshift. \citet{Ito2022} have examined stacked properties of SFGs and QGs at $0<z<5$ of a total number of 322,743. They have found that QGs have higher BHAR/SFRs than SFGs, suggesting that AGN activity plays a crucial role in quenching at high redshifts. 

In these previous  
studies, BHAR/SFRs 
have been
found to be 
lower than expected from the local mass relation, indicating that the growth of $M_\mathrm{BH}$ in inactive galaxies at the probed redshifts is slower than the prediction of simultaneous co-evolution. However, these studies have primarily focused on the relationship between BHAR and $\Mstar$, without directly comparing and discussing the BHAR and SFR in relation to the local correlation. Their BHAR/SFR values may also be biased because they use stellar-mass limited samples.

In this study, we use a large ($N \simeq 12,000$) Lyman-break galaxy (LBG) sample for stacking analysis.
This sample is rest-frame FUV continuum magnitude limited, and hence, approximately SFR limited. 
In addition to SFR, we also examine the relation with the dark halo mass ($\Mh$), or equivalently, the halo accretion rate (HAR), of hosting galaxies using clustering-based dark halo mass estimates available for the sample. Dark halo mass is a primary parameter that determines the evolution of both SMBHs and stellar components \citep{Benson2010,Vogelsberger2020};
BHAR/$\Mh$ (or BHAR/HAR) and SFR/$\Mh$ (SFR/HAR) are measures of their mass-growth efficiencies in dark haloes.
By comparing BHAR/SFR and BHAR/HAR ratios with the corresponding local mass relations, we 
constrain the evolution of SMBHs and host galaxies at earlier cosmic times than probed by the previous studies \footnote{The mean redshifts of sources in \citet{Fornasini2018}'s and \citet{Ito2022}'s highest-redshift bins are both $z \simeq 3.6$.}.

We describe our galaxy sample in Section~\ref{sec: Data}. The X-ray stacking analysis and BHAR calculation methods are described in Section~\ref{sec: Analysis}. Results are given in Section~\ref{sec: results}, where we also compare the results with QSOs and with cosmological simulations of co-evolution. 
We discuss 
several implications of the results in Section~\ref{sec: discussion}.
A summary is given in Section~\ref{sec: summary}.
Throughout this paper, we use a \citet{Chabrier2003} initial mass
function (IMF) 
and assume a flat cosmology with $H_\mathrm{0} = 70$ km s$^{-1}$ Mpc$^{-1}$, $\Omega_\mathrm{\Lambda} = 0.7$, and $\Omega_\mathrm{M} = 0.3$. 
All magnitude are presented in the AB system \citep{Oke1983}.

\section{Data}
\label{sec: Data} 

\subsection{Star-forming Galaxies Sample}
\label{sec: Star-forming Galaxies sample}

We use the catalog of 12,128 LBGs at $4 \lesssim z \lesssim 7$ constructed by \citet{Harikane2022}, considering them to be SFGs. These LBGs have been selected by the standard color selection technique \citep[e.g.][]{Steidel1996, Giavalisco2002} from $g,r,i,z$, and $y$-band data of the UD-COSMOS field taken in the Subaru/Hyper Suprime-Cam Subaru Strategic Program (HSC SSP) \citep{Aihara2019}.

We derive the SFR
for each object as follows \citep{Harikane2022}. First, 
we calculate the UV absolute magnitude ($M_\mathrm{UV}$) from the observed magnitude ``convolvedflux\_0\_20\_mag'' ($i$-band for $z\sim 4$, $z$-band
for $z\sim 5$, and $y$-band
for $z\sim 6,7$) in \citet{Harikane2022}. Then,
we estimate the dust attenuation corrected UV luminosity ($L_{\mathrm{UV}}$) from 
$M_\mathrm{UV}$ using the dust attenuation--UV slope relation by \citet{Meurer1999} 
($A_{1600}=4.43+1.99\beta$, where $A_{1600}$ is the dust attenuation in the rest frame 
$1600$ \AA\ and $\beta$ is the UV slope), 
where we estimate 
$\beta$ from the UV-slope--$L_{\mathrm{UV}}$ relation shown in \citet{Bouwens2014}. Finally, we convert $L_{\mathrm{UV}}$ into SFR using the formula given in \citet{Madau2014} (but after rescaling for a Chabrier IMF): ${\rm SFR} \ [M_\odot \ {\rm yr}^{-1}] = 7.2\times 10^{-29}
L_{\rm UV} \ [{\rm erg \ s^{-1}\ Hz^{-1}}]$.
\citet{Harikane2022} also provide the average 
$M_{\mathrm{h}}$
as a function of
$M_{\mathrm{UV}}$
estimated by clustering analysis 
in their eq.54.

\renewcommand{\arraystretch}{1.3}
\begin{table*}
        \begin{threeparttable}
        \caption{
 LBG samples and sub-samples used in this study.
 }
 	\label{tab:subsample_for_stack}
	\begin{tabular}{ccccccccc}
	\hline
	redshift & 
 original
 $^{a}$ & no 
 Chandra
 $^{b}$ & parent sample $^{c}$ & $M_{\rm UV} \le -23$ $^{d}$ & mag bin & sub-sample $^{e}$ & 
        used for stacking $^{f}$\\
	\hline
	\multirow{5}{*}{$z \sim 4$} & \multirow{5}{*}       {9231 (18)} & \multirow{5}{*}{238}
 & \multirow{5}{*}{8993 (18)} & \multirow{5}{*}{12 (4)} & 23-24 & 95 (2) & 93 \\ 
        & & & & & 24-25 & 1055 (1) & 1054 \\
        & & & & & 25-26 & 4445 (6) & 4439 \\
        & & & & & 26-27 & 3386 (5) & 3381 \\
        & & & & & all & 8981 (14) & 8967 \\ \hline
        \multirow{5}{*}{$z \sim 5$} & \multirow{5}{*}{2665 (2)} & \multirow{5}{*}{62} 
        & \multirow{5}{*}{2603 (2)} & \multirow{5}{*}{1 (0)} & 23-24 & 8 (1) & 7\\ 
        & & & & & 24-25 & 270 (0) & 270 \\
        & & & & & 25-26 & 1488 (0) & 1488 \\
        & & & & & 26-27 & 836 (1) & 835 \\
        & & & & & all & 2602 (2) & 2600 \\ \hline
		$z \sim 6$ & 205 (0) & 4 & 201 (0) & 0 & all & 201 (0) & 201\\ 
  \hline
            $z \sim 7$ & 27 (0) & 0 & 27 (0) & 0 & all & 27 (0) & 27\\ 
		\hline
	\end{tabular}

\textit{Notes.}
 
$^a$ Total number of LBGs in the original catalog.

$^b$ Number of LBGs
outside the Chandra image.

$^c$ Number of LBGs
within the Chandra image.

$^d$ Number of LBGs
brighter than $M_{\rm UV} = -23$.

$^e$ Number of LBGs in each sub-sample.

$^f$ Number of LBGs in each sub-sample used for X-ray stacking.

$^{a,c,d,e}$ The number in
parentheses
indicates the number of individual X-ray detected 
LBGs.

\end{threeparttable}
\end{table*}

\subsection{Chandra X-ray Image}

\begin{figure*}
    \centering
    \begin{subfigure}[b]{1\columnwidth}
        \centering
        \includegraphics[width=\textwidth]{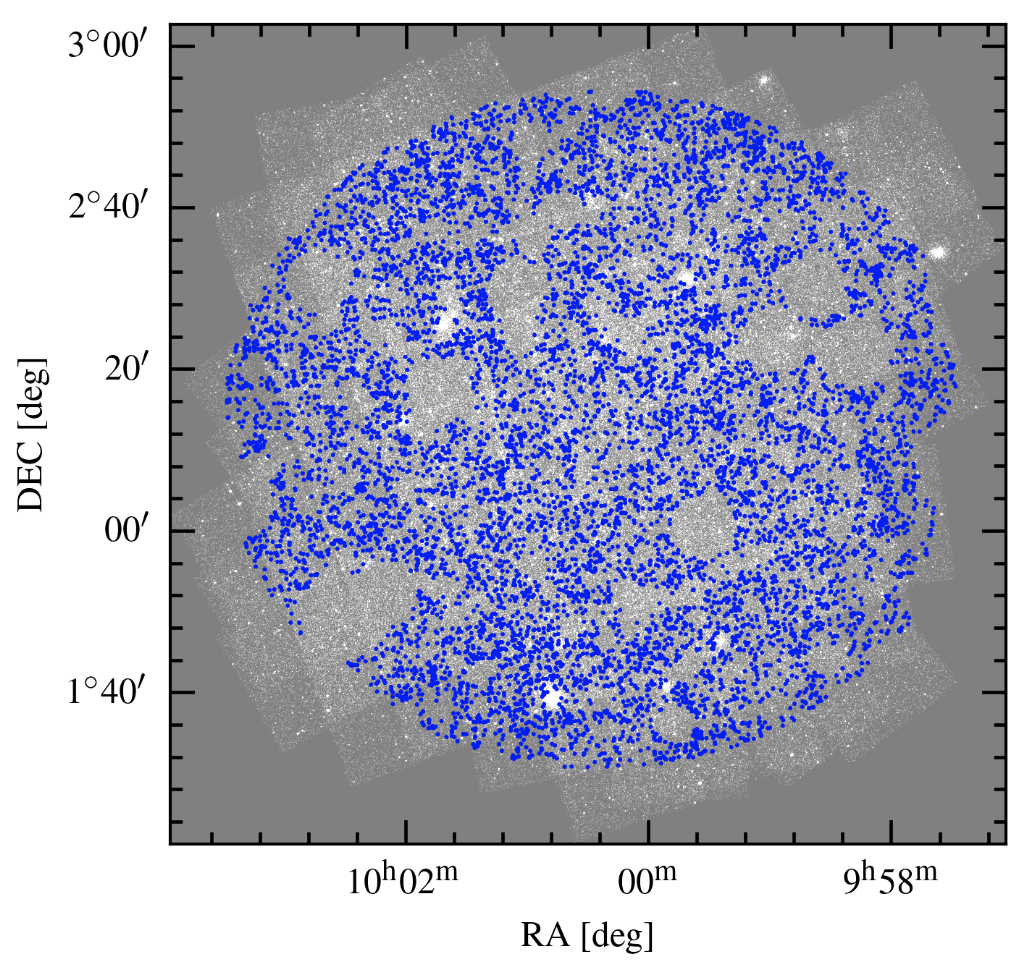}
        \caption{$z\sim4$}
        \label{fig:subfig1}
    \end{subfigure}
    \hfill
    \begin{subfigure}[b]{1\columnwidth}
        \centering
        \includegraphics[width=\textwidth]{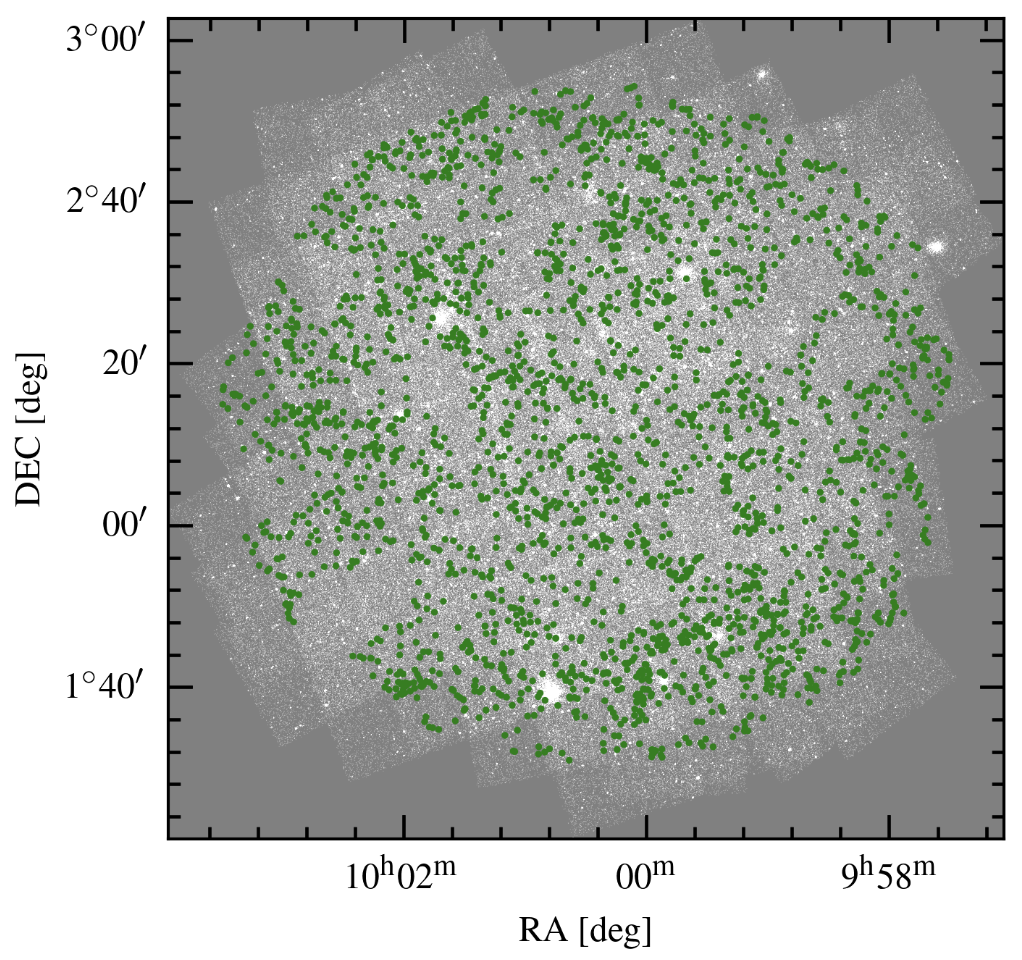}
        \caption{$z\sim5$}
        \label{fig:subfig2}
    \end{subfigure}

    \vspace{0.5cm} 
    
    \begin{subfigure}[b]{1\columnwidth}
        \centering
        \includegraphics[width=\textwidth]{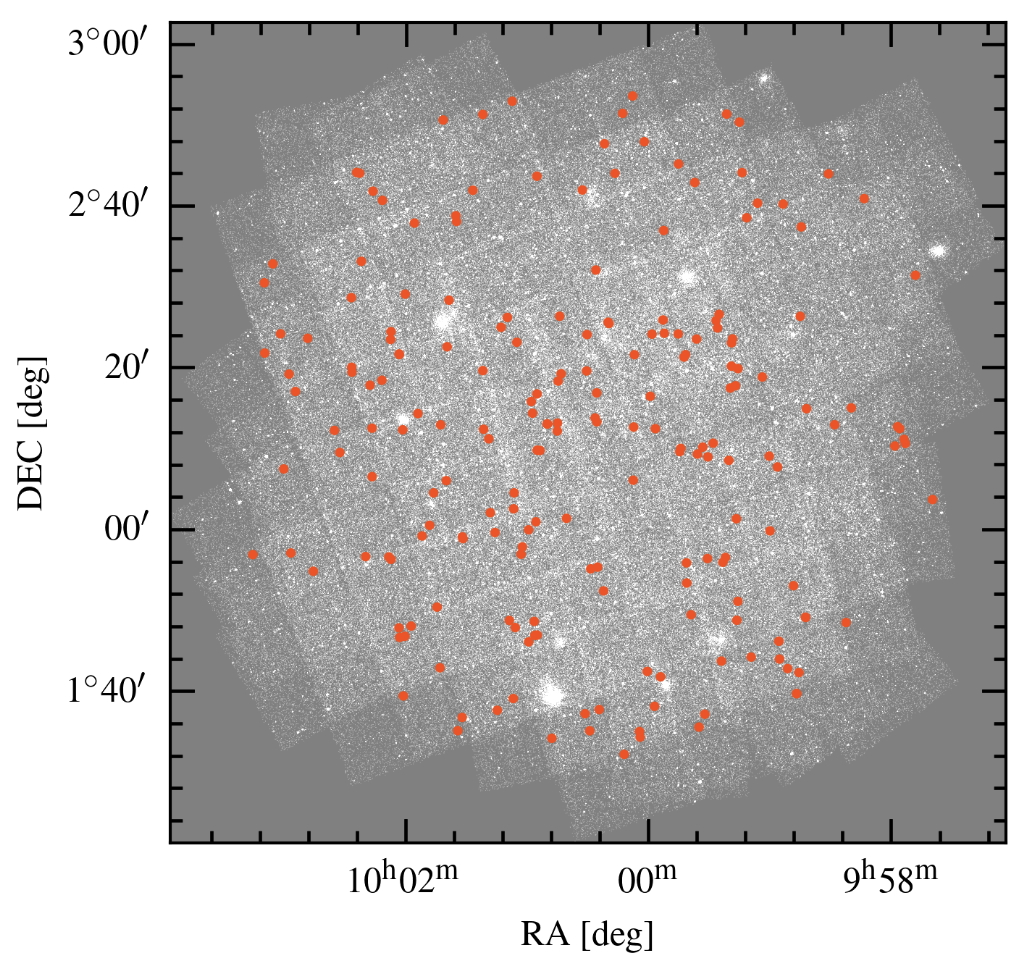}
        \caption{$z\sim6$}
        \label{fig:subfig3}
    \end{subfigure}
    \hfill
    \begin{subfigure}[b]{1\columnwidth}
        \centering
        \includegraphics[width=\textwidth]{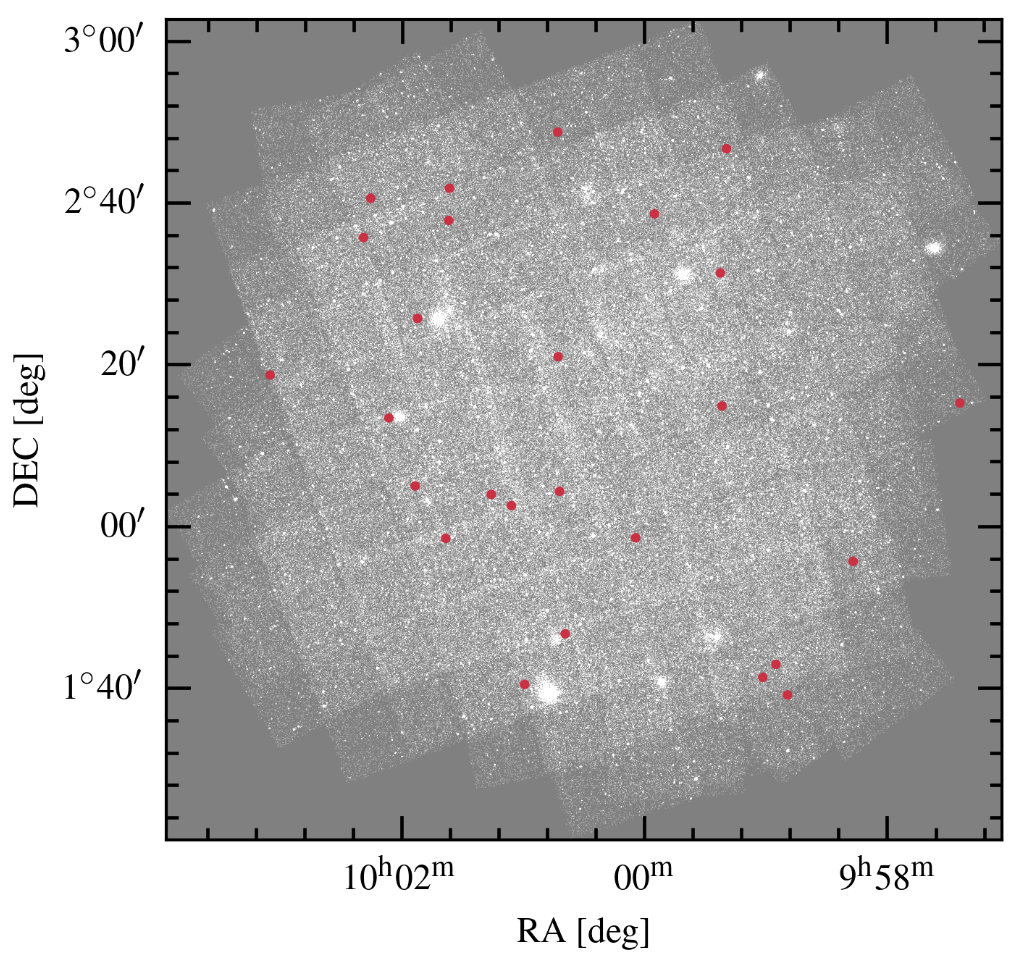}
        \caption{$z\sim7$}
        \label{fig:subfig4}
    \end{subfigure}
    \caption{
    Distributions of LBGs at $z\sim 4,5,6,$ and 7 (blue, green, orange, and red 
    dots, respectively) on the 0.5--2 keV Chandra image. Only those in the parent samples (see Table~\ref{tab:subsample_for_stack}) are plotted.}
    \label{fig:sky_distribution}
\end{figure*}

For X-ray stacking, we use the data from the Chandra COSMOS Legacy survey \citep{Civano2016}, which is a 4.6 Ms Chandra program that has imaged $2.2$ deg$^2$ of the COSMOS field. 
The depth of the data reaches $2.2 \times 10^{-16}$ $\  \mathrm{erg}$ $\mathrm{cm}^{-2}$ $\mathrm{s}^{-1}$ in the 0.5--2 keV band and $1.5 \times 10^{-15}$ $\mathrm{erg}$ $\mathrm{cm}^{-2}$ $\mathrm{s}^{-1}$ in the 2--10 keV band.
Fig.~\ref{fig:sky_distribution} shows the distribution of our LBGs (parent samples in Table.~\ref{tab:subsample_for_stack})
on the 0.5--2 keV image. 

\subsection{X-ray Stacking of Sub-samples}

This study aims at constraining the average BH accretion properties of normal 
SFGs. Even a tiny portion of X-ray detected sources in a sample for stacking can greatly affect the average trend. By cross-matching the LBG catalog with the Chandra COSMOS Legacy Survey source catalog \citep{Civano2016} with a matching radius of $2''$, we find that 18 out of 9231 LBGs at
$z\sim 4$ and two out of 2665 LBGs at $z\sim 5$ are detected while finding no detected source at $z\sim 6$ or 7. We exclude these 20 sources from our X-ray stacking. Their 2--10~keV luminosities after hydrogen absorption correction 
in a similar manner to stacked samples (see Section~\ref{sec: Analysis})
are distributed over 
$1.45 \times 10^{44} - 9.11 \times 10^{44}$ erg~s$^{-1}$,
corresponding to BHARs of $0.37 - 2.31$ $M_{\odot}$ yr$^{-1}$.

We also
exclude sources brighter than $M_{\mathrm{UV}} = -23$.
This is because the UV spectrum of such bright sources may be significantly contaminated by an AGN \citep{Harikane2022}, making $L_{\mathrm{UV}}$-based SFR estimation unreliable. 
X-ray stacking is performed separately for $z\sim 4$, 5, 6, and 7.
For $z\sim 4$ and 5, we also stack magnitude-divided sub-samples with a 1 mag bin. 
Table~\ref{tab:subsample_for_stack} summarises 
the number of sources used for X-ray stacking as well as those excluded. 
This paper defines these Chandra X-ray undetected galaxies as `inactive galaxies'.
They
constitute the vast majority of the galaxy population at the redshifts in question.

\section{X-Ray Stacking Analysis}
\label{sec: Analysis} 

\subsection{Stacking Procedure}
\label{sec: stacking_procedure}

We use the Chandra stacking tool \texttt{CSTACK} v.4.4 \footnote{\href{https://lambic.astrosen.unam.mx/cstack_v4.4/}{https://lambic.astrosen.unam.mx/cstack\_v4.4/}} \citep{Miyaji2008}. 
\texttt{CSTACK} provides an exposure-time weighted average count rate (with error) at the positions of objects given in an input source catalog after excluding objects that are $>8'$ from the optical axis or are affected by resolved X-ray sources. 
An average count rate (CR),
which is defined so that $90\%$ of the total counts are included if stacked objects are PSF-like, is calculated for the soft (0.5--2 keV) and hard (2--8 keV) bands separately. The error in the average 
CR is calculated using bootstrap resampling. \texttt{CSTACK} also generates a stacked $30'' \times 30''$ image for each band. See, e.g., \citet{Ito2022} for a more detailed explanation of \texttt{CSTACK} and an application to inactive high-redshift galaxies.

\begin{figure}

\includegraphics[width=\columnwidth]
{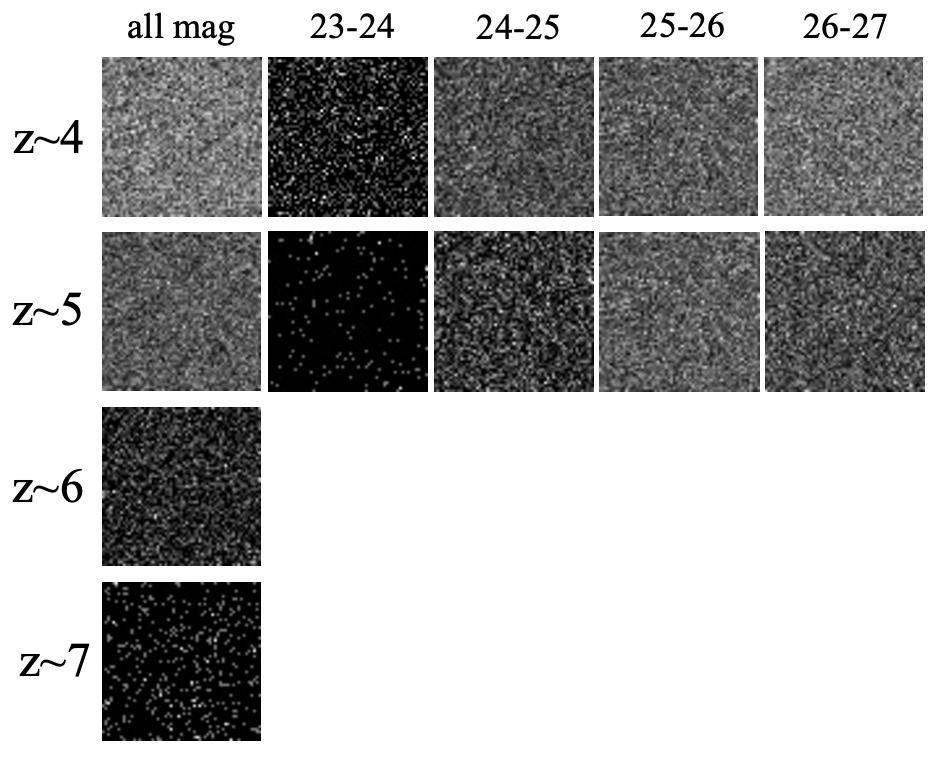}
\caption{Stacked soft-band
X-ray images of our LBG samples.
No signal is detected in any sample. 
The size of each image is $30'' \times 30''$.}
    \label{fig:stacking_result}
\end{figure}

We apply \texttt{CSTACK} to the whole sample at each redshift as well as the magnitude-divided sub-samples at $z\sim 4$ and 5.
No significant ($\ge 3\sigma$) signal is detected in any sample (Fig.~\ref{fig:stacking_result}; only results for the soft band are shown).
Therefore, for each sample, we use a value three times the 
CR error ($\langle \mathrm{CR}_{\mathrm{err}}^{3\sigma} \rangle$) as the $3\sigma$ upper limit
of the average 
CR of the sources.
Because the soft band gives a stronger upper limit to the BHAR 
than the hard band, we use the results from the soft band in this paper.

\subsection{Calculation of Average BHAR Upper Limits}
\label{sec:Calculating_BHAR}
For each sample, we calculate the average BHAR upper limit ($\BHAR$) from $\langle{\mathrm{CR}_{\mathrm{err}}^{3\sigma}}\rangle$ using the following procedure. 
The $\BHAR$ and other quantities of each sample are summarized in Table~\ref{tab:stacked_properties}.

\begin{enumerate}
    \item First, we convert $\langle{\mathrm{CR}_{\mathrm{err}}^{3\sigma}}\rangle$ to 
    the $3\sigma$ upper limit of hydrogen absorption-corrected average
    flux ($\langle {F_{\mathrm{X}}^{3\sigma}} \rangle$) 
    in the soft band using  \texttt{PIMMS}\footnote{\href{https://cxc.harvard.edu/toolkit/pimms.jsp}{https://cxc.harvard.edu/toolkit/pimms.jsp}}. 
    We adopt a photon index of $\Gamma=1.8$ \citep[e.g.][]{Marchesi2016, Ricci2017} for model spectra and consider both the Galactic and internal absorptions.
    We adopt the auxiliary response file 
    of cycle 14. For the hydrogen column density ($\NH$) of the Milky Way, we adopt $\log (\NH/ {\rm cm^{-2}}) =20.4$  obtained by \citet{Kalberla2005}. For the $\NH$ of LBGs, we use an empirical relation of the median $\NH$ for massive ($\Mstar > 10^{10} M_\odot$) galaxies given in \citet{Gilli2022}: $\NH(z)=10^{21.0} (1+z)^{3.3}$ cm$^{-2}$. This relation has been obtained from ALMA data of cold gas for massive galaxies 
    at $0.4 \lesssim
    z \lesssim 6$ and QSOs at $z\sim 6$ and has been found to be consistent with the $\NH$ values of AGNs from X-ray spectroscopy. This relation gives $\log (\NH [z=4] / {\rm cm^{-2}}) =23.2$, $\log (\NH[z=5] / {\rm cm^{-2}}) =23.5$, $\log (\NH[z=6] / {\rm cm^{-2}}) =23.8$, and $\log (\NH [z=7] / {\rm cm^{-2}}) =24.0$, being roughly in line with the average values of the X-ray individually detected sources in our LBG sample derived from X-ray hardness ratios (HR),
    $\log (\NH[{\rm HR}, z\sim4] / {\rm cm^{-2}}) =22.9$, $\log (\NH [{\rm HR}, z\sim5] / {\rm cm^{-2}}) =23.9$, and the average value of $z\sim2-5$ SFGs estimated in \citet{Ito2022}, $\log (\NH[{\rm HR}] / {\rm cm^{-2}}) =23.3$.
    \item Then, we convert
    $\langle {F_{\mathrm{X}}^{3\sigma}} \rangle$ to  
    the absorption-corrected 
    average
    X-ray luminosity in the rest frame 2--10 keV (${\langle L_{2-10 \mathrm{keV}}^{3\sigma}}\rangle$) by: 
    \begin{equation}
    \langle{L_{2-10 \mathrm{keV}}^{3\sigma}}\rangle = \frac{4\pi {{d_{\mathrm{L}}}^2({10}^{2-\Gamma}-{2}^{2-\Gamma})}}{(1+\langle z \rangle)^{2-\Gamma}({E_2}^{2-\Gamma}-{E_1}^{2-\Gamma})} \langle{F_{\mathrm{X}}^{3\sigma}}\rangle, 
	\label{eq:1}
    \end{equation}
    where $d_{\mathrm{L}}$ is the luminosity distance at the average redshift ($\langle z \rangle$) of the sample, and $E_1$ and $E_2$ are the soft-band range (i.e., $E_1=0.5$ keV and $E_2=2$ keV). 
    We find that the $\Lxm$ of each sample is 1.5--2.0 times larger than that before internal
    $\NH$ absorption correction.
    \item 
    X-ray binaries (XRBs)
    can significantly contribute to the 
    $\Lxm$ of inactive galaxies. We remove the contribution of XRBs from 
    $\Lxm$ using an empirical relation given in
    \citet{Lehmer2010}: 
    \begin{align}
    \begin{split}
        L_{\mathrm{X,XRB}} &= 10^{29.37 \pm 0.15}(1+\langle z \rangle)^{2.0 \pm 0.6} \langle \Mstar \rangle \\
        &\quad + \: 10^{39.28 \pm 0.05} (1+\langle z \rangle)^{1.3 \pm 0.1}\mathrm{\langle SFR \rangle}, 
    	\label{eq:2}
    \end{split}
    \end{align}
    where $\langle \Mstar\rangle$ and $\langle $SFR$\rangle$ are the average 
    $\Mstar$
    and SFR of the sample, respectively, and obtain the 
    average
    AGN luminosity, $\langle{L_{\mathrm{X,AGN}}^{3\sigma}}\rangle$, as: 
    \begin{equation}
        \langle{L_{\mathrm{X,AGN}}^{3\sigma}}\rangle = \langle{L_{2-10 \mathrm{keV}}^{3\sigma}}\rangle - L_{\mathrm{X,XRB}}
    	\label{eq:3}
    \end{equation}
    
    We find that the calculated $\LXXRB$ is a factor of $2.6-22$ lower than $\LXAGN$, implying that our results described in the next section are robust against the uncertainty in the 
    XRB correction. We also note that using other empirical relations gives similar results.

    \item 
    We then derive the average bolometric luminosity upper limit ($\langle L_{\mathrm{bol}}^{3\sigma}\rangle$) by multiplying $\langle{L_{\mathrm{X,AGN}}^{3\sigma}}\rangle$ by the conversion factor, $k_{\mathrm{bol}}$ ($\equiv L_{\mathrm{bol}}/L_{\mathrm{X}}$):
    \begin{equation}
        \langle{L_{\mathrm{bol,AGN}}^{3\sigma}}\rangle = k_{\mathrm{bol}} \times \langle{L_{\mathrm{X, AGN}}^{3\sigma}}\rangle.
    \label{eq:4}
    \end{equation}
    \citet{Yang2018} provide $k_{\mathrm{bol}}(L_{\mathrm{X}})$ for $\log (L_{\mathrm{X}} / {\rm erg\ s^{-1}}) > 42.4$ sources, which is a modified version of \citet{Lusso2012}'s and is an increasing function of $L_{\mathrm{X}}$. With this relation, the values of $k_{\mathrm{bol}}$ at our samples' average
    $3\sigma$ upper limits ($\log (L_{\mathrm{X,AGN}}^{3\sigma} / {\rm erg\ s^{-1}}) = 41.85-43.66$) are $\simeq 10 - 13$. On the other hand, \citet{She2017} adopt $k_{\mathrm{bol}} = 16$ for $\log (L_{\mathrm{X}} / {\rm erg\ s^{-1}}) < 42.4$. \citet{Carraro2020} and \citet{Ito2022} use \citet{Yang2018}'s relation for sources brighter than $\log (L_{\mathrm{X}} / {\rm erg\ s^{-1}}) = 42.4$ and $k_{\mathrm{bol}} = 16$ for those fainter.
    In this study, however, we conservatively adopt $k_{\mathrm{bol}}=16$ for all samples. \citet{Fornasini2018} also use this value.
    \item Finally, 
    we calculate 
    $\BHAR$ as:
    \begin{equation}
        \BHAR = \frac{(1-\epsilon) \times {\langle L_{\mathrm{bol, AGN}}^{3\sigma}}\rangle}{\epsilon c^2},
    \label{eq:5}
    \end{equation}
    where $c$ is the speed of 
    light and $\epsilon$ is the radiation efficiency of SMBH mass accretion. 
    We adopt $\epsilon=0.1$ following the previous X-ray stacking studies \citep[e.g.][]{Fornasini2018, Carraro2020, Ito2022}. 
    
\end{enumerate} 

\subsection{X-ray Individual Detection Sources}
\label{sec:individual_detection}
We also calculate the BHAR for the X-ray individual detection sources. There are 18 X-ray detection sources in the $z\sim4$ bin and two sources in the $z\sim5$ bin. 
Among them,
four sources in the $z\sim4$ bin and one source in the $z\sim5$ bin have spectroscopic redshifts 
within their LBG selection windows,
five sources in the $z\sim4$ bin have photometric redshifts 
within the selection windows,
and the remaining sources have 
redshifts outside the selection windows.
We only derive the BHAR for the ten sources within the selection windows.

\clearpage

\begin{center}

\begin{sidewaystable}
    \centering
    \captionsetup{justification=raggedright, singlelinecheck=false}
    \renewcommand{\arraystretch}{1.3}

    \begin{threeparttable}
        \caption{Stacked X-Ray Properties of LBGs.}
        \label{tab:stacked_properties}

    \begin{tabular} 
    {ccccccccccccc}
    	
	\hline
	\multirow{2}{*}{ID} & 
        \multirow{2}{*}{${\langle z \rangle}^{a}$} &
        \multirow{2}{*}{mag bin} & 
        $\langle \mathrm{SFR}\rangle$ $^{b}$ & 
        $\log {\langle \Mstar \rangle}$ $^{c}$ & 
        $\log {\langle \Mh \rangle}$ $^{d}$ &
        $\log {\HAR}$ $^{e}$ &
        $\log {\langle{L_\mathrm{2-10keV}^{3\sigma}}\rangle}$ $^{f}$ & 
        $\log {\langle L_{\mathrm{x,AGN}}^{3\sigma}\rangle}$ $^{g}$ & 
        $\log {\langle{L_{\mathrm{bol,AGN}}^{3\sigma}}\rangle}$ $^{h}$ & 
        $\log {\BHAR}$ $^{i}$\\
         & & & [$\Msun$ yr$^{-1}$] &
         [$\Msun$] & 
         [$\Msun$] &
         [$\Msun$ yr$^{-1}$] &
         [erg s$^{-1}$] &
         [erg s$^{-1}$] &
         [erg s$^{-1}$] &
         [$\Msun$ yr$^{-1}$]\\
		\hline
		0 & 3.8 & (23,24) & 90.2 & 10.53 & 12.96 & 4.21 &
        42.86 & 42.76 & 43.97 & $-$1.83
        \\ 
		1 & 3.8 & (24,25) & 29.9 & 10.06 & 12.19 & 3.37 &
        42.31 & 42.19 & 43.40 & $-$2.40
        \\ 
		2 & 3.8 & (25,26) & 10.8 &  9.61 & 11.67 & 2.79 &  
        42.00 & 41.92 & 43.12 & $-$2.68
        \\ 
            3 & 3.8 & (26,27) & 5.1  &  9.28 & 11.40 & 2.49 &
            42.06 & 42.03 & 43.23 & $-$2.57
        \\ 
            4 & 3.8 &   all   &  11.7 &  9.65 & 11.78 & 2.94 & 
            41.85 & 41.71 & 42.92 & $-$2.89
        \\ 
            5 & 4.9 & (23,24) &  124.8 & 10.68 & 12.67 & 4.11 &
            43.66 & 43.63 & 44.84 & $-$0.96
        \\ 
            6 & 4.9 & (24,25) &  
            47.8
            & 10.26 & 12.19 & 3.59 & 
            42.86 & 42.79 & 43.99 & $-$1.81
        \\ 
            7 & 4.9 & (25,26) &  16.4 &  9.79 & 11.72 & 3.07 &
            42.48 & 42.42 & 43.62 & $-$2.18
        \\ 
            8 & 4.9 & (26,27) &  8.1 &  9.49 & 11.47 & 2.79 &
            42.60 & 42.58 & 43.79 & $-$2.01
        \\ 
            9 & 4.9 &   all   &  17.3 &  9.82 & 
            11.76
            & 3.11 &
            42.36 & 42.28 & 43.48 & $-$2.32
        \\ 
           10 & 5.9 &   all   &  35.2 & 10.13 & 11.91 & 3.46 &
           43.15 & 43.12 & 44.32 & $-$1.48
        \\ 
           11 & 6.9 &   all   &  59.7 & 10.36 & 12.05 & 3.76 &
           43.66 & 43.64 & 44.84 & $-$0.95
        \\ \hline
	\end{tabular}
 
\textit{Notes.}

$^a$ Average redshift of the sample.

$^b$ Average SFR of the sample.

$^c$ Average stellar mass
of the sample 
calculated from the average SFR using the SFR--$\Mstar$ relation of main sequence galaxies \citep{Pearson2018}.

$^d$ Average dark halo mass of the sample.

$^e$ Average halo accretion rate 
derived from $\Mhm$ by eq.~\ref{eq:6} in Section~\ref{sec: BHAR_vs_HAR}.

$^f$ Average X-ray luminosity $3\sigma$ upper limit.

$^g$ Average AGN X-ray luminosity $3\sigma$ upper limit.

$^h$ Average AGN bolometric luminosity $3\sigma$ upper limit.

$^i$ Average BHAR $3\sigma$ upper limit.

\end{threeparttable}

\end{sidewaystable}

\end{center}

\clearpage

\section{Results}
\label{sec: results} 

\subsection{BHAR versus SFR}
\label{sec: BHAR_vs_SFR} 

\begin{figure*}
	\includegraphics[width=\textwidth]{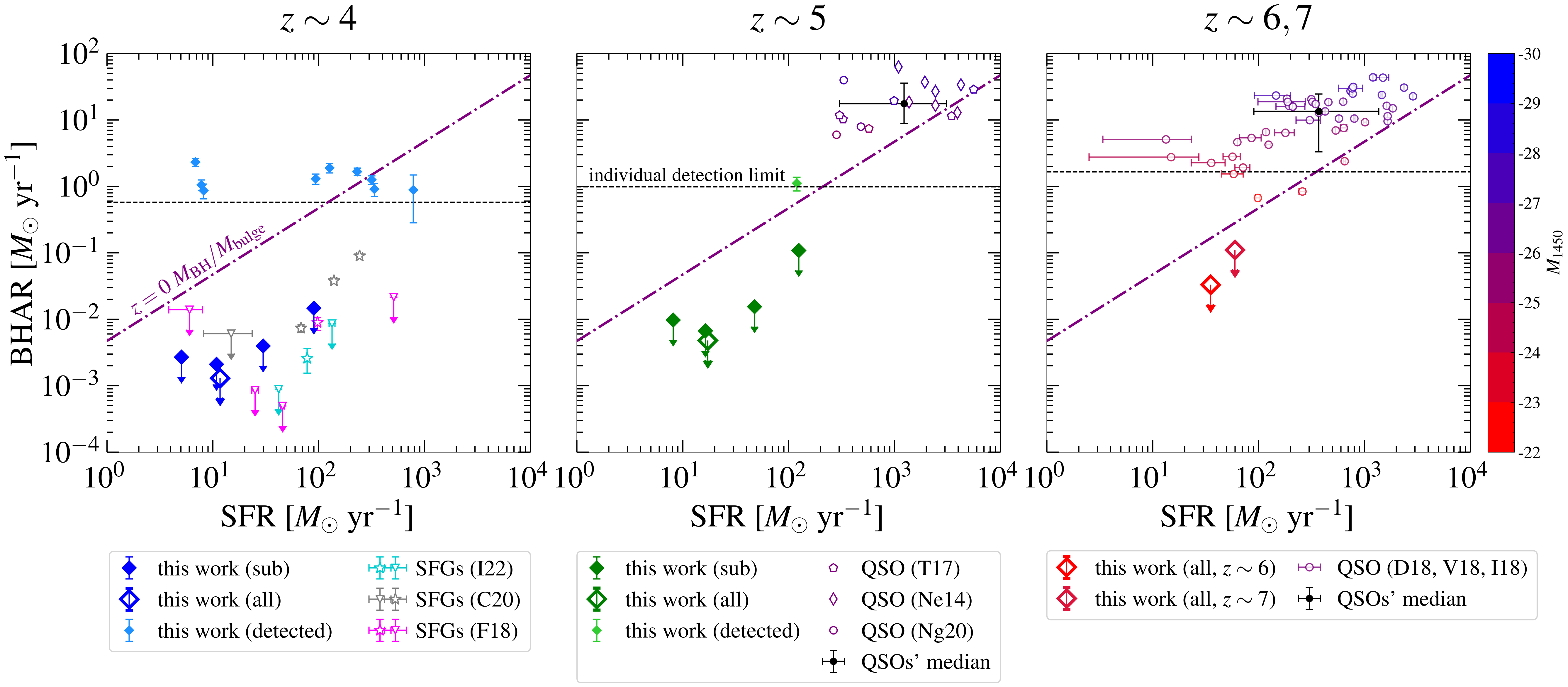}
    \caption{
    BHAR plotted against SFR for $z\sim 4$ (left panel), $z\sim 5$ (middle), and $z\sim 6,7$ (right). Open diamonds with down arrows represent the $3\sigma$ upper limit of all LBGs at each redshift while filled diamonds with down arrows are for magnitude-divided sub-samples. 
    Small open symbols indicate QSOs, color-coded by $M_\mathrm{1450}$ ($z\sim 5$: \citet{Trakhtenbrot2017} with $-25.94 > M_\mathrm{1450} > -27.29$; \citet{Netzer2014} with $-26.11>M_\mathrm{1450}>-27.87$; \citet{Nguyen2020} with $-25.86>M_\mathrm{1450}>-27.62$, $z\sim 6,7$: \citet{Decarli2018,Izumi2018,Venemans2018} with $-22.83>M_\mathrm{1450}>-29.30$). 
    Black dots with error bars indicate the median and the central 68 percentile of the distribution of all QSOs at each redshift. 
    Small filled diamonds in blue or green represent individually X-ray detected LBGs, for which the SFR is tentatively estimated from $\MUV$ for plotting purpose. 
    Also plotted in the left panel for comparison are the stacking results of \citet{Fornasini2018}, \citet{Carraro2020}, and \citet{Ito2022} at slightly lower redshifts than this study (open star symbols if detected and inverted triangles with down arrows for $3\sigma$ upper limits). 
    For consistency, 
    the $2\sigma$ upper limits given in \citet{Ito2022} and \citet{Carraro2020} have been converted to $3\sigma$ upper limits.
    Dot-dashed lines indicate the local $\MBH/\Mbulge$ ratio ($\mathrm{BHAR}/\mathrm{SFR}=4.7\times10^{-3}$) \citep{Kormendy2013} while dashed lines indicating the limit of individual X-ray detection in the Chandra Legacy Survey \citep{Civano2016} at each redshift:
    $\log (L_\mathrm{x}/{\rm erg \ s^{-1}}) \gtrsim 44.12$ at $z\sim4$, $\log (L_\mathrm{x}/{\rm erg \ s^{-1}}) \gtrsim 44.31$ at $z\sim5$, $\log (L_\mathrm{x}/{\rm erg \ s^{-1}}) \gtrsim 44.43$ at $z\sim6$, and $\log (L_\mathrm{x}/{\rm erg \ s^{-1}}) \gtrsim 44.50$ at $z\sim7$.
    }
    \label{fig:BHAR_vs_SFR}
\end{figure*}

In Fig.~\ref{fig:BHAR_vs_SFR}, 
we plot ${\BHAR}$ against ${\langle \mathrm{SFR}\rangle}$ for our stacked samples together with QSOs at similar redshifts in the literature.
The dot-dashed line in each panel corresponds to the local $M_\mathrm{BH}/M_\mathrm{bulge}$ ratio 
calculated from the data of early-type galaxies given in \citet{Kormendy2013} (their Tables 2 and 3), after correction for the mass return fraction ($R=0.41$ for a Chabrier IMF), which is the fraction of stellar mass that is ejected back to the interstellar medium (e.g., \citet{Udeda2018}).
If SMBHs and their host galaxies are growing in tandem, they are on this line with $\mathrm{BHAR}/\mathrm{SFR}=(1-R)(M_\mathrm{BH}/M_\mathrm{bulge})_{\rm local} = 4.7 \times 10^{-3}$, where $(M_\mathrm{BH}/M_\mathrm{bulge})_{\rm local}=8.0\times10^{-3}$.
Here we have assumed that at $z=0$, our LBGs have evolved into early-type galaxies with $M_\mathrm{bulge} \approx \Mstar$ and used galaxies with $\Mh \ge 1 \times 10^{13} M_\odot$ for the calculation.
This assumption is reasonable
because the present-day descendants of our LBGs ($11.4 < \log (\Mhm/ M_\odot) <13.0$) are expected to have $\Mh \gtrsim 1\times 10^{13} M_\odot$ \citep[e.g.][]{Behroozi2013}, at which masses most of the galaxies hosted are quenched \citep{Behroozi2019}. 
The ratio is not sensitive to the $\Mh$ range used for the calculation because the local $\MBH$ -- $\Mbulge$ relation by \citet{Kormendy2013} is close to linear ($\MBH \propto \Mbulge^{1.17}$). 

We find that at $z\sim 4$ and 5 the $\BHAR$ of our LBGs are $\sim 1.5$ dex and $\sim 1$ dex lower than the dot-dashed line, respectively, implying that the SMBHs of SFGs at these redshifts are growing much more slowly than expected from simultaneous evolution. At $z\sim 6,7$, the upper limits are only up to $\sim 0.5$ dex below the line, possibly because of the small numbers of stacked sources.

QSOs are plotted at $z\sim 5$ and $6,7$. Those at $z\sim 5$ are bright sources selected from SDSS QSOs while those at $z\sim 6,7$ are discovered by several wide-field optical surveys and include faint sources from the HSC SSP. Their BHARs have been estimated from rest-frame 3000 \AA\ or 1450 \AA\ luminosities in \citet{Netzer2014} and from rest-frame 1450 \AA\ luminosities in \citet{Trakhtenbrot2017, Nguyen2020, Venemans2018} while their SFRs from IR luminosities. We find that the QSOs have $\sim1$ dex higher BHAR/SFR ratios than the $z=0$ mass ratio, implying that the SMBHs in QSOs are growing much faster than expected from simultaneous evolution. As a result, the difference in BHAR/SFR between LBGs and QSOs amounts to $\sim 2$ dex.

In the left panel of Fig.~\ref{fig:BHAR_vs_SFR}, we also plot the results of three previous stacking studies at their highest-redshift bins: $\langle z \rangle = 2.6-2.8$ \citep{Carraro2020}, $\langle z \rangle = 3.2-3.6$ \citep{Fornasini2018}, and $\langle z \rangle = 3.5-3.6$ \citep{Ito2022}. All three use the same Chandra data as this study. \citet{Fornasini2018}'s and \citet{Ito2022}'s estimates and upper limits are lower than our upper limits. This is probably because they can reach fainter luminosity limits at lower redshifts and, in some cases, with larger sub-samples than ours. \citet{Carraro2020}'s results are higher than these two studies, being closer to our upper limits, probably because they have included sources with X-ray individual detection. Note that we cannot accurately compare the results of these three studies with ours because they use stellar-mass limited samples and constructed sub-samples by splitting the parent sample by specific SFR and $\Mstar$, not by SFR.

\subsection{BHAR versus HAR}
\label{sec: BHAR_vs_HAR} 

\begin{figure*}
	\includegraphics[width=\textwidth]{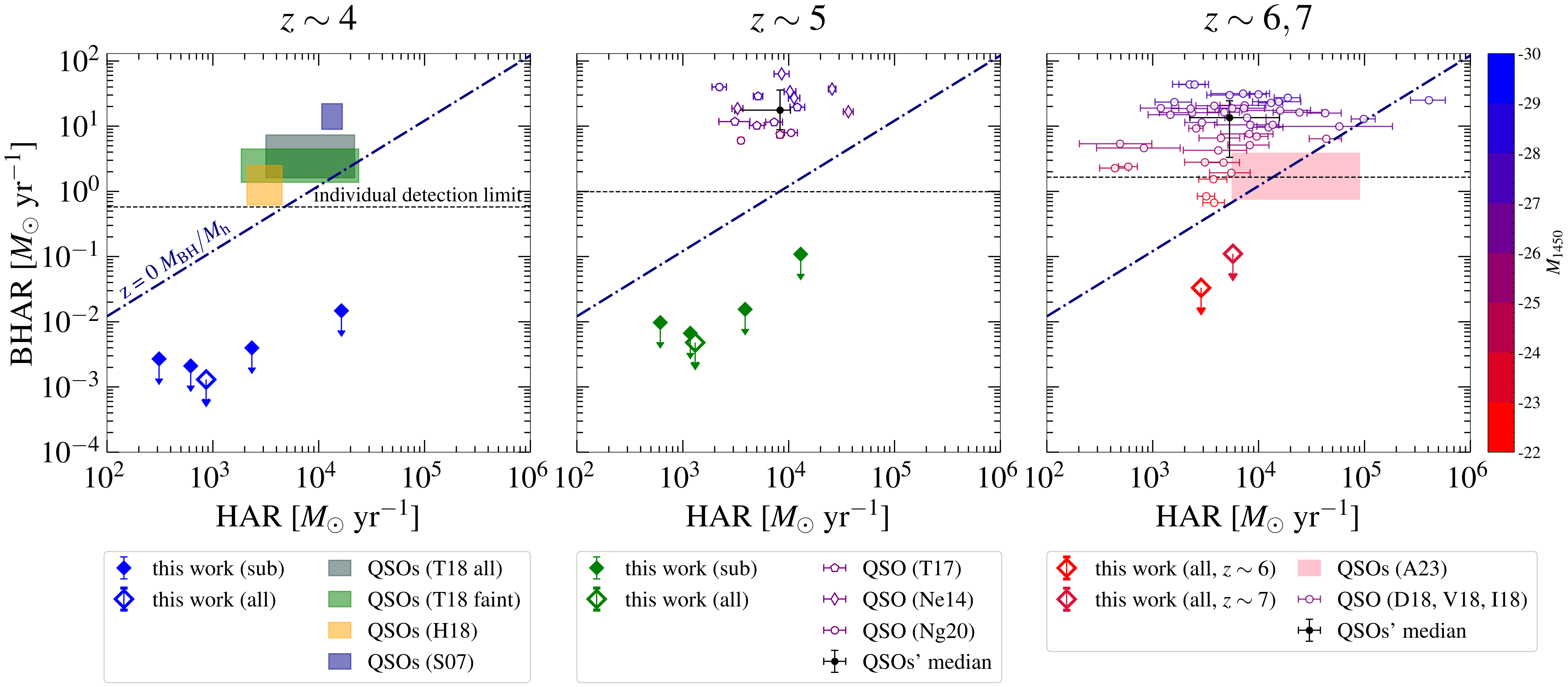}
    \caption{
    BHAR plotted against HAR for $z\sim 4$ (left panel), $z\sim 5$ (middle), and $z\sim 6,7$ (right). The meaning of 
    diamonds with down allows,
    small open symbols, black dots with error bars, and 
    dashed lines is the same as in Fig~\ref{fig:BHAR_vs_SFR}. The $\Mh$ of individual QSOs, which are needed to calculate HAR, are taken from (or calculated in the same manner as) \citet{Shimasaku2019}. Shaded areas on the left and right panels correspond to QSO samples with clustering-based $\Mh$ estimates \citep{Timlin2018,He2018,Shen2007,Arita2023}; 
    for a given QSO sample, the 
    range of the HAR corresponds to the uncertainty in 
    $\Mh$ estimate while the 
    range of the BHAR 
    corresponds to the 
    central 68 percentile of the $M_\mathrm{1450}$ distribution,
    where we have calculated $\Lbol$ from $M_\mathrm{1450}$ using 
    eq.1 of \citet{Venemans2016}.
    The median $M_\mathrm{1450}$ of these QSO samples are 
    $M_\mathrm{1450} = -24.58$ (all) and 
    $M_\mathrm{1450} = -24.39$ (faint) \citep{Timlin2018}, 
    $M_\mathrm{1450} = -23.49$ \citep{He2018}, 
    $M_\mathrm{1450} = -26.35$
    \citep{Shen2007},
    and $M_\mathrm{1450} = -23.92$ \citep{Arita2023}. 
    Dot-dashed lines indicate the local $\MBH/\Mh$ ratio ($\mathrm{BHAR}/\mathrm{HAR}=1.2\times10^{-4}$)
    \citep{Kormendy2013}.
    }
    \label{fig:BHAR_vs_HAR}
\end{figure*}

We then compare the mass growth rate of SMBHs with that of hosting dark haloes (HAR: 
halo accretion rage) in Fig.~\ref{fig:BHAR_vs_HAR}.
We estimate the $\HAR$ of each LBG sample from its $\Mhm$ using the "mean" version of \citet{Fakhouri2010}'s eq.2:
\begin{align}
    \begin{split}
        \HAR &= 46.1M_{\odot} \ \mathrm{yr^{-1}}\left(\frac{\Mhm}{10^{12}M_{\odot}}\right)^{1.1} \\
        &\quad \times (1+1.11\z)\sqrt{\Omega_\mathrm{M}(1+\z)^3+\Omega_\mathrm{\Lambda}}, 
    \end{split}
    \label{eq:6}
\end{align}
which has been obtained from
Millennium and Millennium-II simulation results \citep{Springel2005, Boylan-Kolchin2009}. 
The $\Mhm$ of each sample is calculated from the $\MUV-\Mh$ relation in \citet{Harikane2022}.
The dot-dashed line in each panel corresponds to the local $M_\mathrm{BH}/M_\mathrm{h}$ ratio ($\mathrm{BHAR}/\mathrm{HAR}=(1-R)(\MBH/\Mh)_\mathrm{local}=1.2\times 10^{-4}$).
We obtain this ratio using the data of local galaxies given in \citet{Kormendy2013} (their Tables 2 and 3), limiting to those with $\Mh \ge 1\times 10^{13} M_\odot$ for the same reason as for $\MBH/\Mbulge$.

We find that at $z\sim 4$, the LBGs' $\BHAR/\HAR$ is 
$\sim 1-2$ dex
lower than the $z=0$ mass ratio. The difference becomes smaller at $z\sim 5$ and higher, but the LBGs' $\BHAR/\HAR$ is still $\sim 1$ dex lower at $z\sim 5$ and $\sim 0.5$ dex lower at $z\sim 6,7$. These results are qualitatively similar to those found for $\BHAR$ versus $\SFR$.

QSOs from the literature are also plotted in each panel. 
Small open symbols indicate individual QSOs; the $\Mh$ of  \citet{Trakhtenbrot2017,Netzer2014,Nguyen2020}'s QSOs are calculated in the same manner as \citet{Shimasaku2019} and those of \citet{Decarli2018}'s, \citet{Izumi2018}'s, and
\citet{Venemans2018}'s QSOs are taken from \citet{Shimasaku2019}. 
Shaded areas correspond to QSO samples with clustering-based $\Mh$ estimates \citep{Shen2007,He2018,Timlin2018}.
We find that the QSOs are mostly above the $z=0$ mass ratio 
except for \citet{Arita2023}'s  sample at $z\sim 6$,
with individual objects being distributed $\sim 1-2$ dex above the line, 
implying that the SMBHs in QSOs are growing faster than hosting dark matter haloes.
This result appears to be qualitatively in accord with observations that high-redshift QSOs have overmassive SMBHs with respect to the local $\MBH$--$\Mh$ relation \citep{Trainor2012,Shimasaku2019}.
The difference in BHAR/HAR between LBGs and QSOs reaches $\sim 2$ dex or more.

\subsection{SFR versus HAR}
\label{sec: SFR_versus_HAR} 
\begin{figure*}
    \includegraphics[width=\textwidth]{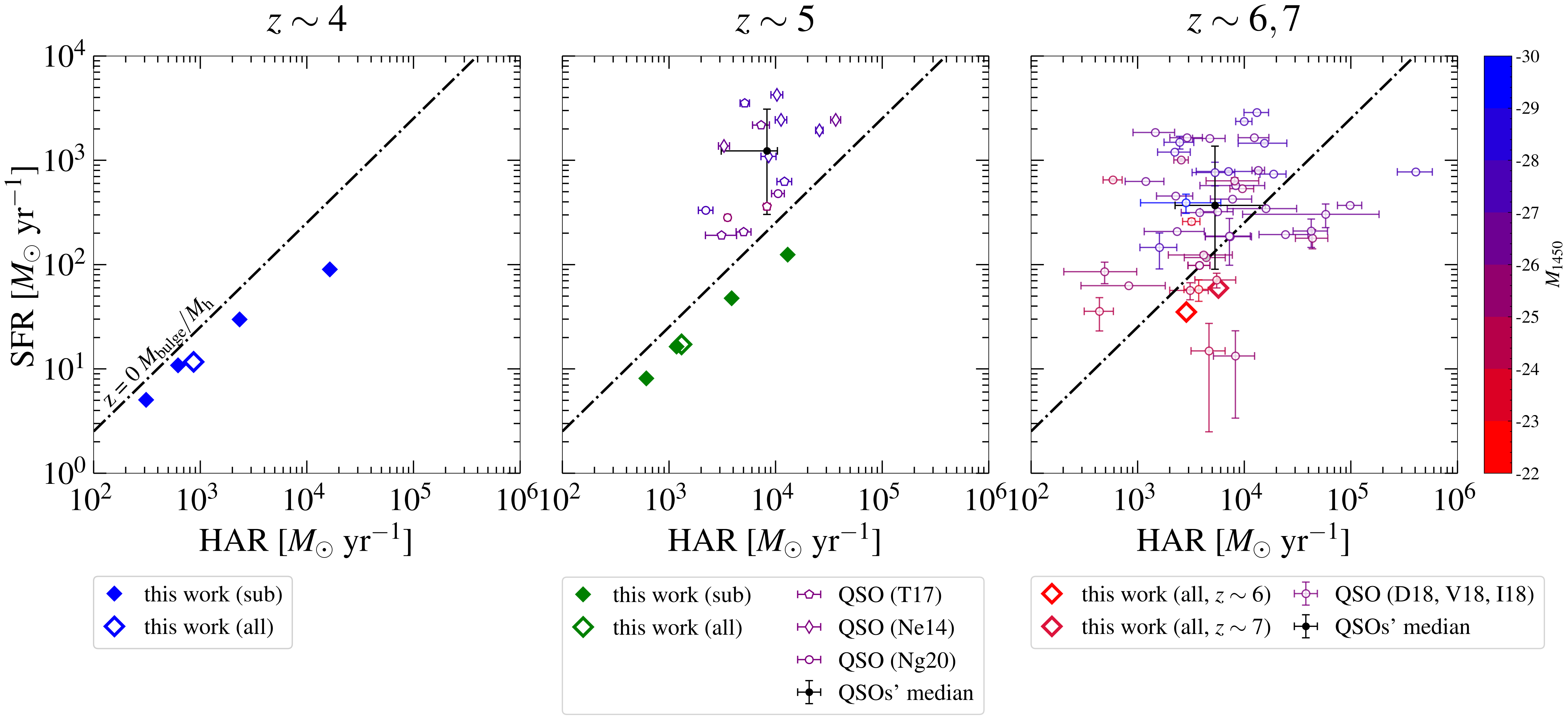}
    \caption{
    SFR plotted against HAR for $z\sim 4$ (left panel), $z\sim 5$ (middle), and $z\sim 6,7$ (right). Open (all objects) and filled (magnitude-divided objects) diamonds indicate LBGs.
    The meaning of small open symbols and black dots with error bars is the same as in Fig~\ref{fig:BHAR_vs_SFR}.
    Dot-dashed lines show the local $\Mbulge/\Mh$ ratio ($\mathrm{SFR}/\mathrm{HAR}=2.5\times10^{-2}$)
    \citep{Kormendy2013}.
    }
    \label{fig:SFR_vs_HAR}
\end{figure*}

We also examine SFR versus HAR for LBGs and QSOs in Fig.~\ref{fig:SFR_vs_HAR}.
We convert the $z=0$ $\Mbulge/\Mh$ ratio \citep{Kormendy2013} to SFR/HAR $=(1-R)(\Mbulge/\Mh)_\mathrm{local}=2.5\times10^{-2}$ in a similar manner to BHAR/SFR, 
where $(\Mbulge/\Mh)_\mathrm{local}=1.5\times10^{-2}$.
We find that the $\SFR/\HAR$ of LBGs 
agree with 
this ratio 
within $\sim 0.3$ dex 
except for the most massive bin at $z\sim 4$.
This means that in SFGs at these redshifts the stellar component and the dark halo are growing roughly in tandem ({\lq}co-evolving{\rq}) on average, in contrast to the stellar component and the SMBH.
On the other hand, most QSOs have higher SFR/HAR than the local mass ratio. The median SFR/HAR of $z\sim 5$ ($z\sim 6,7$) QSOs is found to be $\sim 1$ dex ($\sim 0.5$ dex) higher than the local mass ratio.

\subsection{Comparison with Large-scale Cosmological Simulations}
\label{sec: simu} 

We compare our results with
the TNG300, TNG100 \citep{Marinacci2018,Naiman2018,Nelson2018,Pillepich2018,Springel2018}, SIMBA100 \citep{Dave2019}, and EAGLE100 \citep{Crain2015,Schaye2015,McAlpine2016}
simulations, all of which are large-scale cosmological hydrodynamical simulations and are open to the public.
The TNG300 and TNG100 assume $\epsilon = 0.2$, whereas the SIMBA100 and EAGLE100 adopt $\epsilon = 0.1$. 
Because our $\BHAR$ values are for $\epsilon = 0.1$, we 
recalculate them with $\epsilon=0.2$
when comparing with TNG300 and TNG100. 
For the accretion model, all but SIMBA adopt the Bondi-Hoyle-Lyttleton formalism or a modified version of it, while SIMBA uses a torque-limited accretion model for cold gas and the Bondi-Hoyle-Lyttleton model for hot gas. For a brief summary of these simulations 
(e.g., box size, spatial and mass resolutions, BH seed mass, model of accretion and feedback),
see, e.g., \citet{Habouzit2021}.

To select inactive galaxies from each simulation, we only use 
galaxies below the Chandra X-ray detected limits: $\log (L_\mathrm{x} / {\rm erg \ s^{-1}}) \lesssim 44.12$ at $z\sim4$, $\log (L_\mathrm{x} / {\rm erg \ s^{-1}}) \lesssim 44.31$ at $z\sim5$, $\log (L_\mathrm{x} / {\rm erg \ s^{-1}}) \lesssim 44.43$ at $z\sim6$, and $\log (L_\mathrm{x} / {\rm erg \ s^{-1}}) \lesssim 44.50$ at $z\sim7$.
To imitate the observation, we use 
the SFR averaged 
over 100 Myr for 
TNG300, TNG100 \citep[see][]{Valentino2020}, and SIMBA100. For TNG300 and TNG100, we adopt the stellar mass within twice
the stellar half-mass radius.
We only use SFR $>1 M_\odot \ {\rm yr}^{-1}$ sources in TNG300, TNG100, and EAGLE100, and SFR $>10 M_\odot \ {\rm yr}^{-1}$ in SIMBA100 to ensure the completeness.
 
We adopt the group catalogs of 
Snapshots 21 ($z=4.01$), 17 ($z=5.00$), and 12 ($z=6.49$) in TNG300 and TNG100, the galaxy catalogs of Snapshots 051 ($z=3.96$), 042 ($z=5.02$), and 033 ($z=6.46$) in SIMBA100, and the galaxy catalogs of Snapshots 10 ($z=3.98$), 8 ($z=5.04$), 6 ($z=5.97$), and 5 ($z=7.05$) 
in EAGLE100 to match the redshift bins of our LBG samples.

\subsubsection{Comparing BHAR versus SFR with simulations}
\label{sec: simu_BHAR_SFR}

\begin{figure*}
	
    \begin{subfigure}[b]{1\textwidth}
        \centering
        \includegraphics[width=\textwidth]{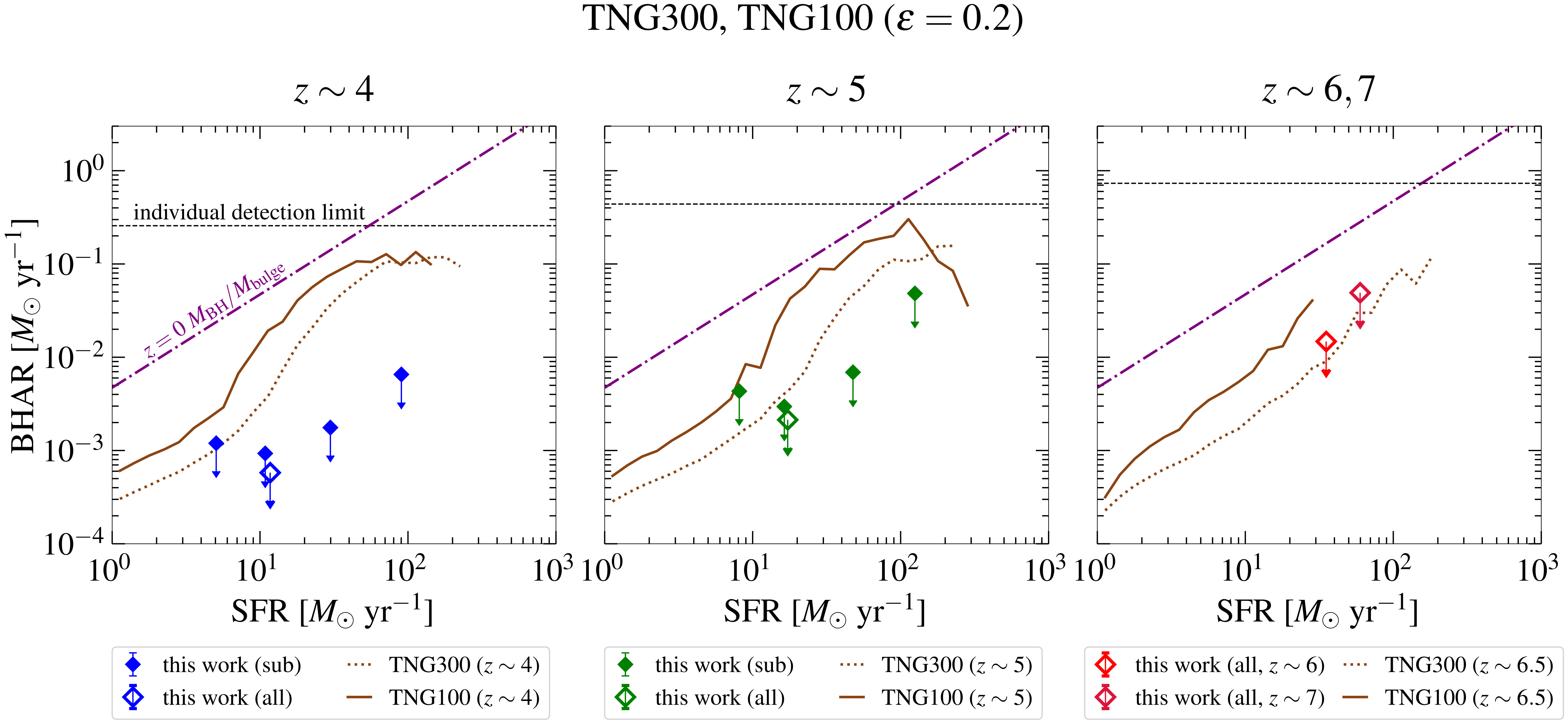}
        \label{fig:first_sub}
    \end{subfigure}
    
    \vspace{0.5cm}

    \begin{subfigure}[b]{1\textwidth}
        \centering
        \includegraphics[width=\textwidth]{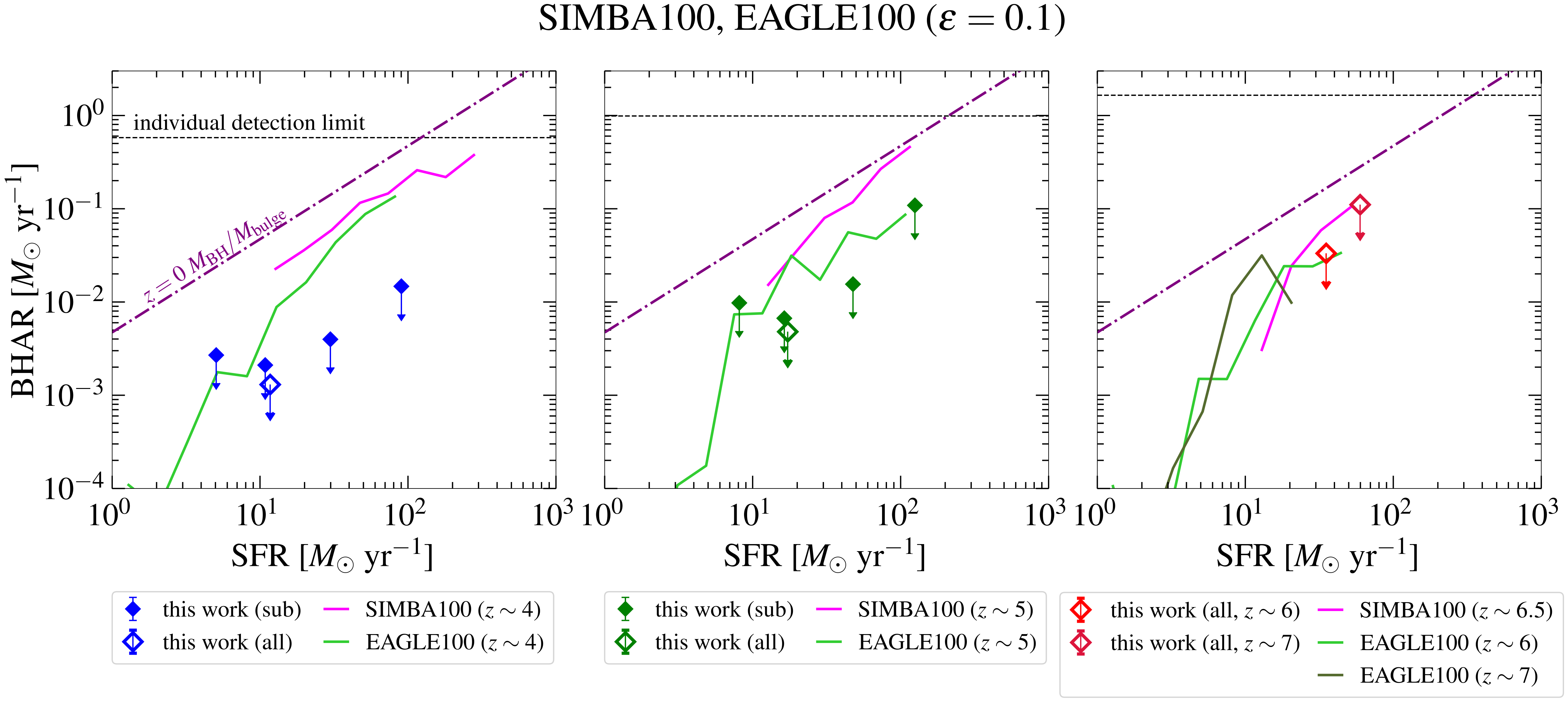}
        \label{fig:second_sub}
    \end{subfigure}
    
\caption{
{\it Top panels.} 
MeanBHAR as a function of SFR predicted by TNG300 (brown dotted lines) and TNG100 (brown solid lines) with $\epsilon=0.2$ for $z\sim4$ (left panel), 5 (middle), and 6.5 (right). {\it Bottom panels.} Same as the top panels but for SIMBA100 (magenta lines) and EAGLE100 (limegreen lines at $z\sim 4,5,6$ and a green line at $z\sim 7$) with $\epsilon=0.1$. Since EAGLE100 does not provide a $z=6.5$ snapshot, we plot $z=6$ and 7 results. When calculating these functions, we only use 
SFR $>1 M_\odot \ {\rm yr}^{-1}$
sources for TNG300, TNG100, and EAGLE, SFR $>10 M_\odot \ {\rm yr}^{-1}$ sources for SIMBA100
to ensure that each simulation is complete in terms of SFR. We examine the completeness using the SFR versus $\Mstar$ plot.
The meaning of diamonds with down arrows and dashed
and dot-dashed lines is the same as in Fig.~\ref{fig:BHAR_vs_SFR}. 
}
\label{fig:simu_BHAR_vs_SFR}
\end{figure*}

First, we compare our $\BHAR$ versus $\SFR$ with the simulations in Fig.~\ref{fig:simu_BHAR_vs_SFR}. As mentioned earlier regarding the difference in $\epsilon$, we divide the plot into two parts: TNG300 and TNG100 in one, and SIMBA100 and EAGLE100 in the other. 

We first look at the results for $z\sim 4$ where our upper limits are most stringent. We find that all four simulations predict much higher BHARs than the data. The largest discrepancy is seen in TNG100, with $\sim 1-1.5$ dex overprediction at SFR $\simeq 10-100 M_\odot$ yr$^{-1}$, but the other three simulations also overshoot the data by $\sim 1$ dex. Note that our data are upper limits, meaning that true differences will be even larger.
The discrepancy is reduced at $z\sim 5$, but TNG100 still predicts $\sim 1$ dex higher values.
At $z\sim 6,7$, all simulations are consistent with our $3\sigma$ upper limits.

\subsubsection{Comparing $SFR$ versus $M_\mathrm{h}$ with Simulations}

\begin{figure*}
    \includegraphics[width=\textwidth]{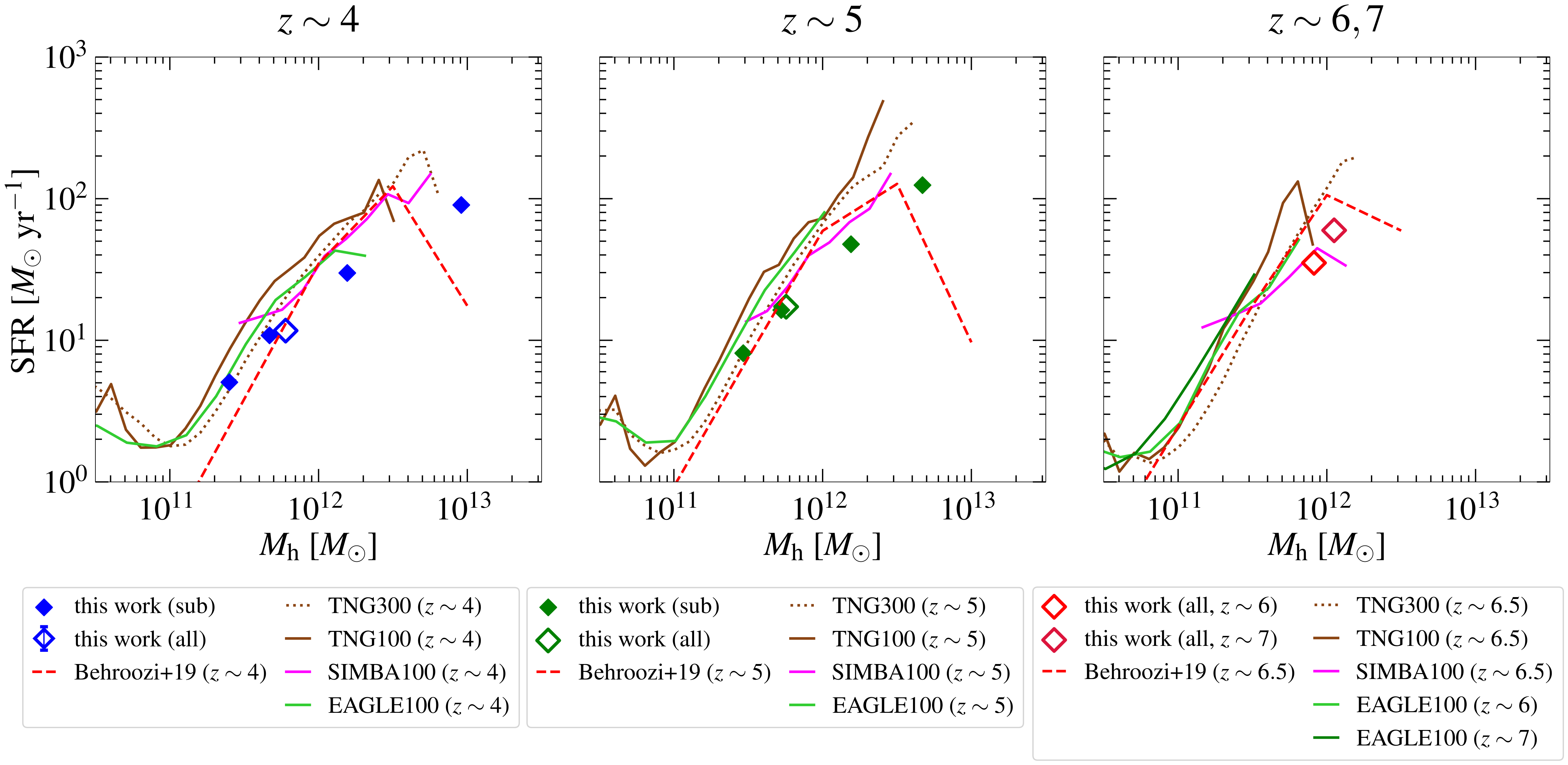}
    \caption{
    Mean SFR as a function of $\Mh$ 
    predicted by the four simulations (TNG300, TNG100, SIMBA100, and EAGLE100) at $z\sim 4$ (left panel), $z\sim 5$ (middle), and $z\sim 6,7$ (right).
    Our LBGs are shown as open diamonds (total samples) and filled diamonds (magnitude divided samples).
    Red dashed lines represent the average relation of all galaxies at each redshift given in \citet{Behroozi2019}.
    }
    \label{fig:SFR_vs_Mh}
\end{figure*}

To constrain the cause of the overprediction found above, we compare the
$\SFR$ versus $\Mhm$ of our LBGs with the simulations in 
Fig.~\ref{fig:SFR_vs_Mh}. 
We find that the simulations correctly predict $\SFR$
as a function of 
$\Mhm$,
with differences from the observations within $\sim 0.3$ dex.
Note also that the simulations are consistent with \citet{Behroozi2019}'s empirical relations.
Therefore, the discrepancy seen in BHAR versus SFR is not due to the star formation physics but the modeling of BH accretion. In the inactive phase, the simulations feed the central SMBH in a given dark halo too efficiently while producing stars in accord with observations.

\section{Discussion}
\label{sec: discussion} 

In the previous section, we find that at $4 \lesssim z \lesssim 7$,
our LBGs’ $\BHAR/\SFR$ and $\BHAR/\HAR$ are 
$\sim 1-1.5$ and $\sim 1-2$ dex
lower than the corresponding $z=0$ mass ratios.
This result leads us to conclude that 
at these redshifts, SFGs cannot obtain as high $M_\mathrm{BH}/M_\mathrm{bulge}$ and $\MBH/\Mh$ as the local mass ratios 
without experiencing a rapid accretion phase (e.g., QSO phase).
In Section~\ref{sec: growth_model}, we compare the contribution of SFGs and QSOs to the growth of SMBHs.
In Section~\ref{sec: SF_AC_property}, we discuss the
star formation and BH
accretion activities of SFGs, QSOs, and two other galaxy populations among which an evolutionary link has been suggested.
We also find that the simulations overpredict the 
inactive galaxies' BHAR
in Section~\ref{sec: simu_BHAR_SFR},
so we discuss 
implications of this result
in Section~\ref{sec: simu-imply}.

\subsection{Contribution of SFGs and QSOs to SMBH Growth}
\label{sec: growth_model}

We compare the contribution of SFGs and QSOs to SMBH growth using our LBG samples and literature QSO samples with duty-cycle ($\fduty$) estimates, 
where $\fduty$ is defined as the fraction of QSO-hosting dark haloes among all haloes over a certain halo mass range. 
Only \citet{He2018}'s and \citet{Shen2007}'s samples at $z\sim4$ have simultaneous estimates of BHAR, $\Mh$ (hence HAR), and $\fduty$. The former sample consists of faint sources with a median of $M_\mathrm{1450} = -23.49$
while the latter consisting of bright sources with 
a median of $M_\mathrm{1450} = -26.35$.
The mean $\Mh$ (with uncertainty)
of these samples, $\log (\Mh/M_\odot) = 12.2-12.5
$ for \citet{He2018} and $\log (\Mh / M_\odot) = 12.8-12.9$ for \citet{Shen2007}, are roughly comparable to those of our $z\sim 4$ LBGs.

\begin{figure}
	\includegraphics[width=\columnwidth]{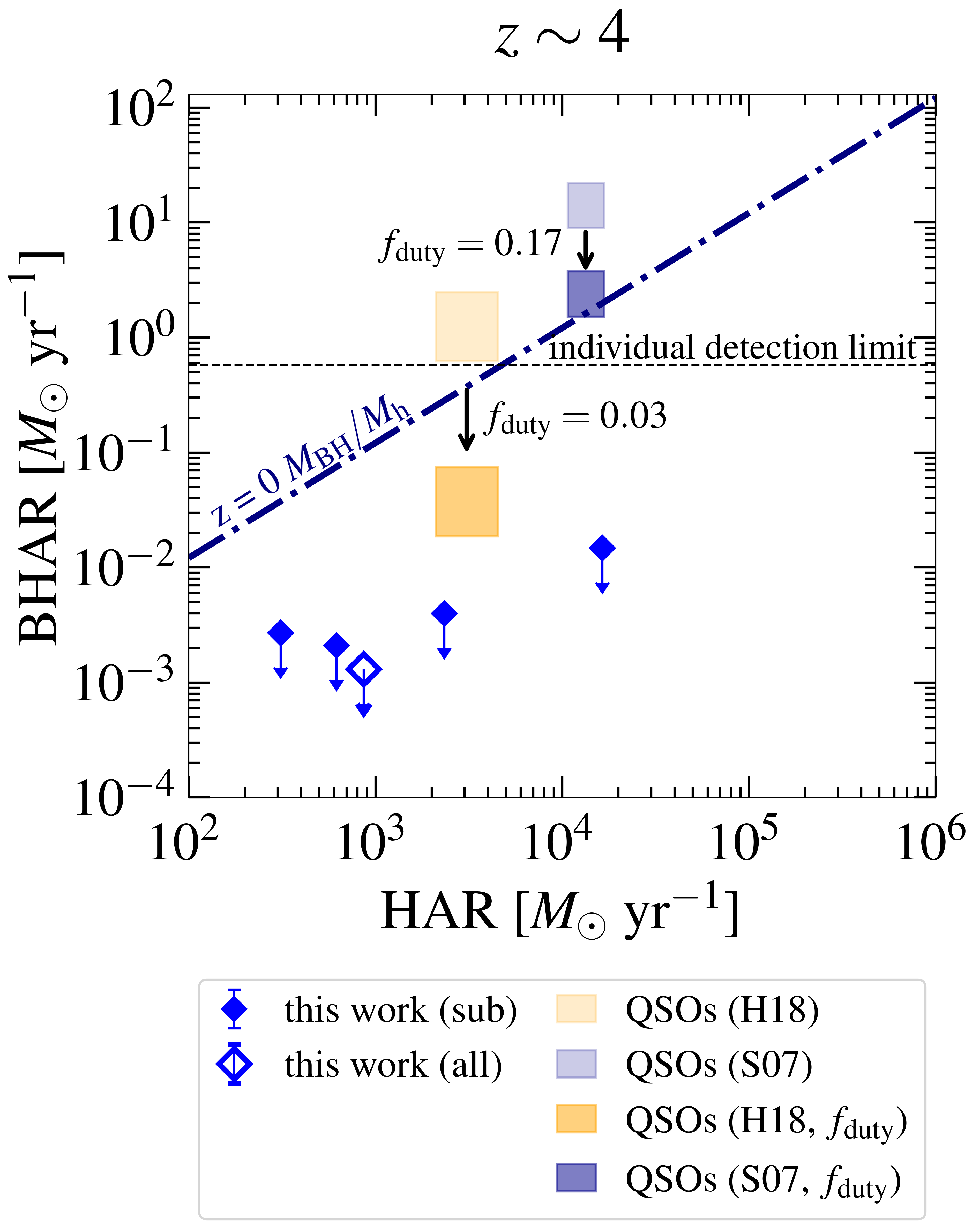}
    \caption{
    BHAR versus HAR for \citet{He2018}'s (yellow shades) and \citet{Shen2007}'s (blue shades) QSO samples at $z\sim 4$ before (light color) and after (dark color) $\fduty$
    correction. The correction is made by multiplying the observed BHAR by
    $\fduty$
    (0.03 for \citet{He2018} and 0.17 for \citet{Shen2007}). Also plotted are our LBGs (diamonds with down arrows) and the local $\MBH/\Mh$ ratio ($\mathrm{BHAR}/\mathrm{HAR}=1.2\times10^{-4}$)
    of \citet{Kormendy2013} (dot-dashed line). 
    The meaning of a dashed line is the same as in Fig.~\ref{fig:BHAR_vs_SFR}.
    }
    \label{fig:BHAR_vs_HAR_DC}
\end{figure}

In Fig.~\ref{fig:BHAR_vs_HAR_DC}, we shift the two QSO samples' original BHAR ranges 
downwards by multiplying them by $\fduty=0.03$ \citep{He2018} and $0.17$ \citep{Shen2007}, respectively, 
to show effective, or time-averaged, BHARs. 
Here, we assume that all dark haloes with 
$\Mh$ similar to those of our LBGs experience a faint QSO phase like \citet{He2018}' QSOs and a bright QSO phase like \citet{Shen2007}'s with a duration indicated by each $\fduty$ estimate.
We find that the effective BHARs are still higher than the 
$\BHAR$ of LBGs with similar HAR by $\sim 0.7$ dex \citep{He2018} and $\sim 2$ dex \citep{Shen2007}. Although the $\fduty$ estimates have large uncertainties ($\sim 1-1.5$ dex [see Fig.11 of \citet{He2018}]), 
the $\MBH$ increase in the QSO phase is likely to dominate over that in the much longer, inactive (BHAR $ \lesssim 1 M_\odot$ yr$^{-1}$) period.
We also find that the effective BHAR/HAR of \citet{Shen2007}'s sample is comparable to the local mass ratio while that of \citet{He2018}'s is much lower.
This implies that if a galaxy 
experiences a bright QSO phase 
like \citet{Shen2007}'s QSOs, 
the mass increase in its SMBH is roughly consistent with simultaneous evolution.

\subsection{Star Formation and BH Accretion Activities of Various Galaxy Populations}
\label{sec: SF_AC_property}

\citet{Hopkins2008} have proposed a co-evolution model that links SFGs, starburst galaxies (SBGs), QSOs, and elliptical galaxies on the assumption that major, gas-rich mergers cause QSO activity.
Regardless of the correctness of this model, it is interesting to quantitatively compare star formation and BH accretion activities between these populations at the redshifts covered by our LBG sample. To do so, we use SFR$/\Mh$ and BHAR$/\Mh$ because both stars and SMBHs grow in their common hosting dark haloes.
We consider LBGs to be normal SFGs in the star-formation main sequence. For SBGs, we use the catalog of IR-detected galaxies in COSMOS provided by \citet{Jin2018}, in which SFRs and AGN luminosities have been estimated by SED fitting to $K_s$ to IR (or radio) photometry. 
They are essentially 
dusty SBGs
with SFR $\simeq 80-2000 M_\odot$ yr$^{-1}$, typically being located 
$\sim 1$ dex 
above the star-formation main sequence.
For ellipticals, we use 
X-ray stacked BHAR estimates for QGs at $z\sim 2-5$ given by \citet{Ito2022} although their average redshift, $z\sim3.2-3.4$,
is slightly lower than ours. We estimate the $M_\mathrm{h}$ of SBGs and QGs from their $\Mstar$ using the conversion formula given by \citet{Behroozi2019}.

\begin{figure*}
    
    \begin{subfigure}[b]{1\textwidth}
        \centering
        \includegraphics[width=0.8\textwidth]{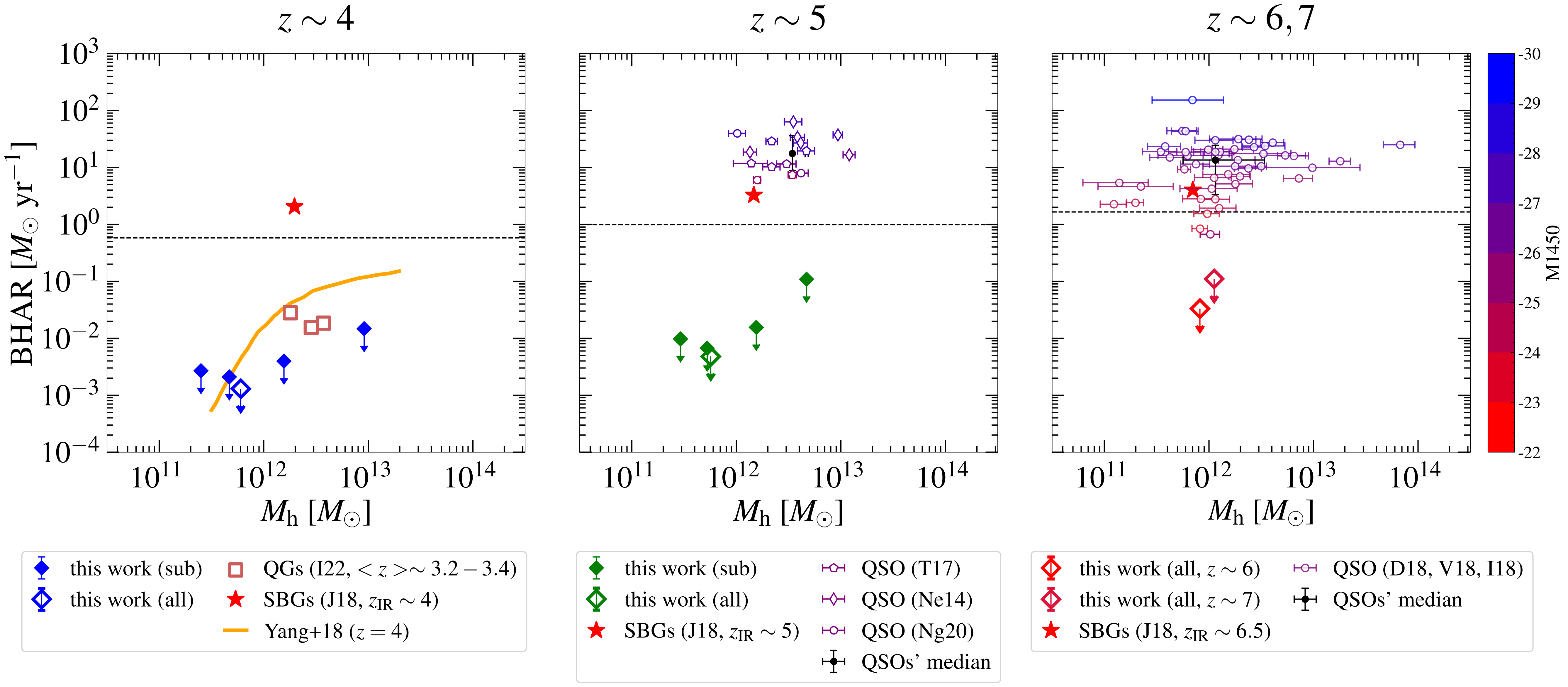}
        \label{fig:BHAR_vs_Mh}
    \end{subfigure}
    
    \vspace{0cm} 

    \begin{subfigure}[b]{1\textwidth}
        \centering
        \includegraphics[width=0.8\textwidth]{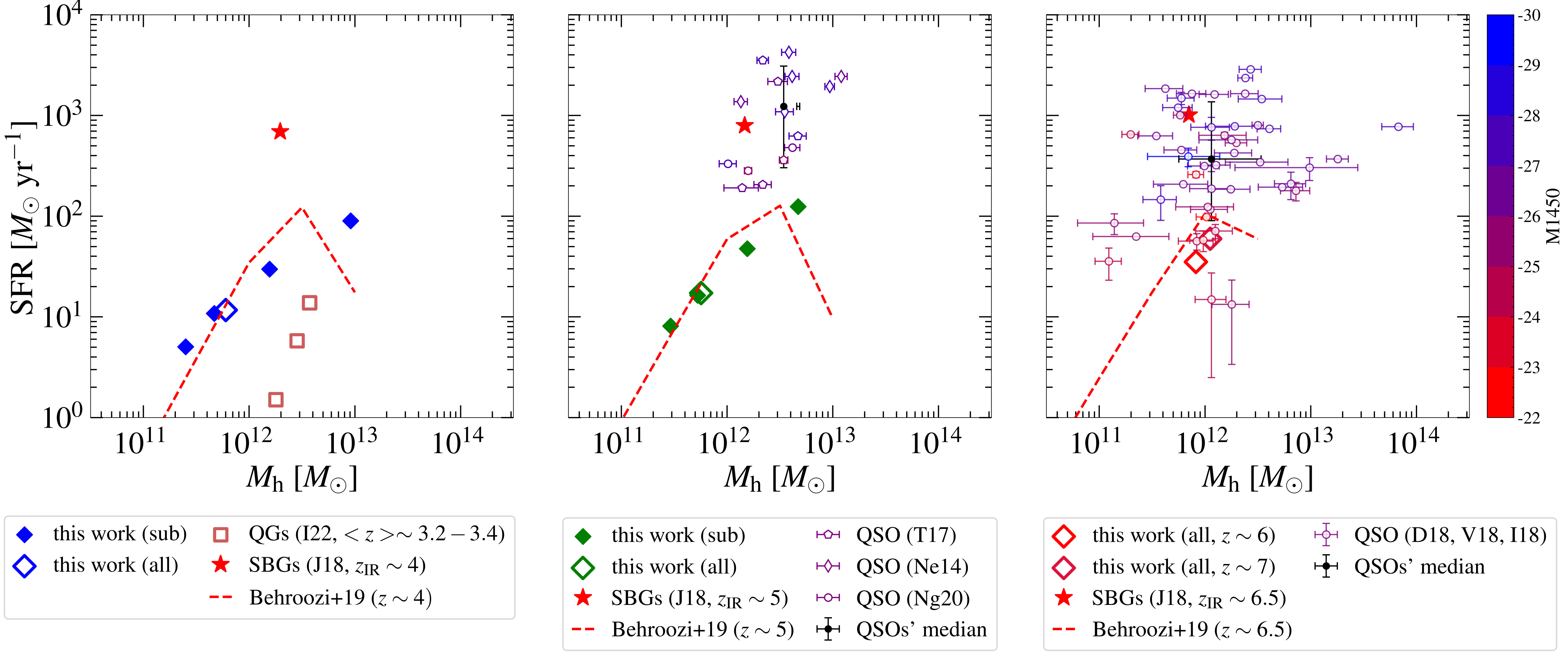}
        \label{fig:SFR_vs_Mh_QSO}
    \end{subfigure}

    \vspace{0cm} 

    \begin{subfigure}[b]{1\textwidth}
        \centering
        \includegraphics[width=\textwidth]{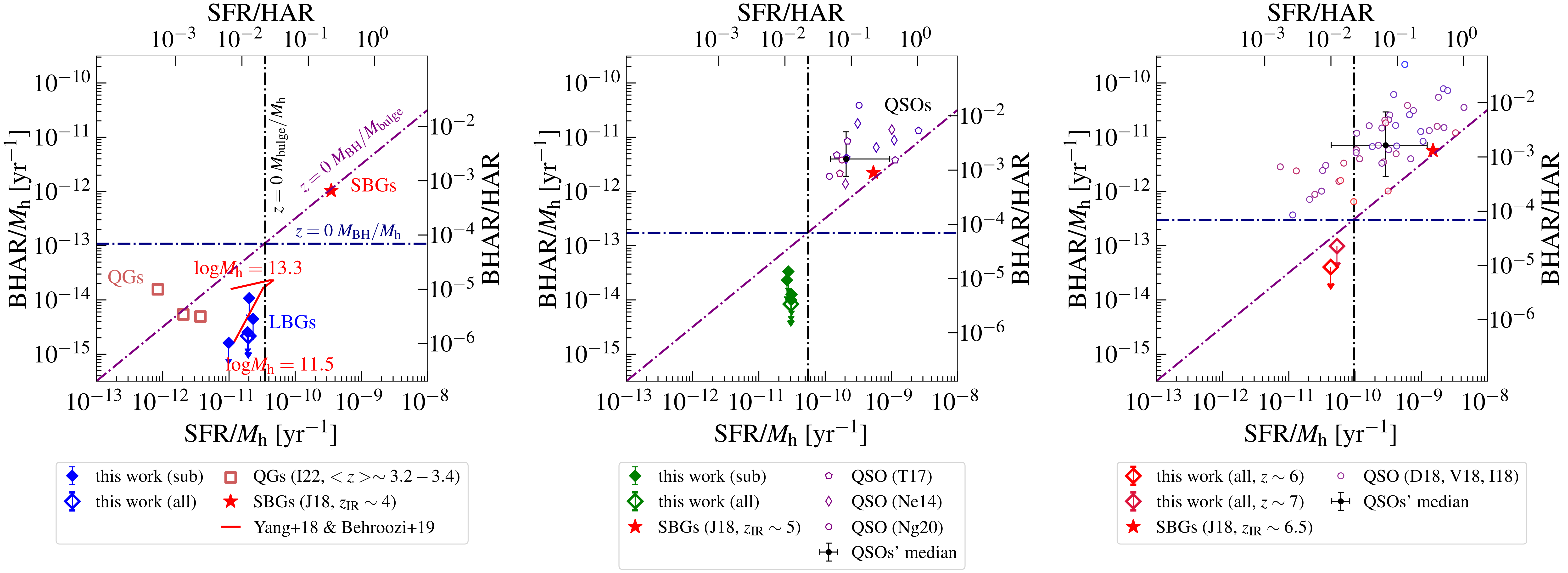}
        \label{fig:BHAR_SFR_Mh_HAR}
    \end{subfigure}
    \caption{{\it Top panels.} BHAR versus $\Mh$ 
    for LBGs (SFGs),
    QSOs (only those with SFR measurements), SBGs, and QGs at three redshift bins. The meaning of diamonds with down arrows, 
    black dots with error bars,
    small open symbols, and 
    dashed lines is the same as in Fig.~\ref{fig:BHAR_vs_SFR}. The newly added orange solid line represents the BHAR versus $\Mh$ relation for all galaxies at $z=4$
    obtained by
    \citet{Yang2018}. Open squares represent QGs 
    at $z\sim3.2-3.4$
    from \citet{Ito2022} and
    filled stars are for SBGs from \citet{Jin2018}. 
    We calculate the $\Mh$ of QGs (SBGs) from their $\Mstar$ using the average $\Mstar-\Mh$ relation for QGs (SFGs) at $z\sim4$ given in \citet{Behroozi2019}.
    {\it Middle panels.} SFR versus $\Mh$ for LBGs, 
    QSOs, SBGs, and QGs at three redshift bins. The meaning of diamonds and red dashed lines is the same as in Fig.~\ref{fig:SFR_vs_Mh}. 
    {\it Bottom panels.} 
    Distributions of LBGs, QSOs, SBGs, and QGs in the BHAR$/\Mh$ -- SFR$/\Mh$ plane at three redshift bins calculated from the top and middle panels.
    In each panel, the upper $x$ axis shows SFR/HAR while the right $y$ axis showing BHAR/HAR; we calculate HAR from $\Mh$ 
    for these two axes
    by approximating 
    the term $\left(\frac{\Mh}{10^{12}M_{\odot}}\right)^{1.1}$ of \citet{Fakhouri2010}'s eq.2 as ${\left(\frac{\Mh}{10^{12}M_{\odot}}\right)}$, i.e., 
    $\langle \mathrm{HAR} \rangle
    \simeq 46.1M_{\odot}\mathrm{yr^{-1}}\left(\frac{\Mhm}{10^{12}M_{\odot}}\right)\times (1+1.11\z)\sqrt{\Omega_\mathrm{M}(1+\z)^3+\Omega_\mathrm{\Lambda}}$, so that HAR is exactly proportional to $\Mh$. This approximation is accurate within 0.1 dex over $\Mh=10^{11} - 10^{13} M_\odot$ in which almost all sources plotted here are included.
    Purple dot-dashed lines indicate the local $\MBH/\Mbulge$ ratio (same as in Fig.~\ref{fig:BHAR_vs_SFR}), navy dot-dashed lines the local $\MBH/\Mh$ ratio (same as in Fig.~\ref{fig:BHAR_vs_HAR}), and black dot-dashed lines the local $\Mbulge/\Mh$ ratio (same as in Fig.~\ref{fig:SFR_vs_HAR}). A solid red line in the left panel  
    shows the average relation of all galaxies over $\log (\Mh/M_\odot) = 11.5-13.3$ obtained by 
    combining the BHAR$-\Mh$ relation \citep{Yang2018} in the top left panel and the SFR$-\Mh$ relation \citep{Behroozi2019}  
    in the middle left panel.
    }
    \label{fig:sBHAR_vs_sSFR}
\end{figure*}

The bottom panels of Fig.~\ref{fig:sBHAR_vs_sSFR} plot BHAR$/\Mh$ against SFR$/\Mh$ for the four populations (LBGs [SFGs],
SBGs, QSOs, and QGs),
obtained from the top 
(BHAR versus $\Mh$)
and middle panels 
(SFR versus $\Mh$).
The QSOs plotted here have $-25.86>M_\mathrm{1450}>-27.87$ at $z\sim5$ \citep{Trakhtenbrot2017,Netzer2014,Nguyen2020} and $-22.83>M_\mathrm{1450}>-29.3$ at $z\sim6,7$ \citep[][and reference therein]{Shimasaku2019}.
The red solid line in the left panel represents the average relation for all $z\sim4$ galaxies with $\Mh = 10^{11.5} - 10^{13.3} M_\odot$ calculated from the empirically obtained SFR versus $M_\mathrm{h}$ relation \citep{Behroozi2019} and BHAR versus $M_\mathrm{h}$ relation \citep{Yang2018}. 
Three dot-dashed lines in each panel correspond to the local $\MBH/\Mbulge$, $\MBH/\Mh$, and $\Mbulge/\Mh$ ratios 
from \citet{Kormendy2013},
respectively. If, for example, a source is located above the $\MBH/\Mh$ line, then the SMBH of this source is growing faster than the dark halo relative to simultaneous evolution.

We find that the QSOs and SBGs are located in a similar region, with $\sim 1$ dex higher SFR$/\Mh$ and $\gtrsim 2$ dex higher BHAR$/\Mh$ than the LBGs, indicating that in QSOs and SBGs, BH accretion activity is more enhanced than star formation activity with respect to SFGs (For the QSOs, we are focusing on the median values of all objects at each redshift).
The $\sim 1$ dex enhancement of SFR$/\Mh$ in QSOs and SBGs compared with LBGs implies that if every SFG experiences a QSO phase and a starburst
phase like plotted here, the 
stellar mass acquired in these phases will dominate over that by normal star formation if the 
$\fduty$ of these phases is higher than $\sim 10\%$.  

A detailed comparison between the QSOs and SBGs shows that, on average, 
the SBGs have slightly higher SFR$/\Mh$ at a fixed BHAR$/\Mh$. This result is consistent with the finding by \citet{Andonie2022}
for $z<3$ infrared QSOs that obscured ($\NH > 10^{22}$ cm$^{-2}$) systems have $\approx 3$ times higher SFRs than unobscured ones with similar 
$\Mstar$.
The BHAR/SFR of our SBGs is comparable to the local $\MBH/\Mbulge$ ratio, meaning that their SMBH and stellar component are growing at paces roughly consistent with simultaneous evolution.

We also find that the QGs have comparable or higher BHAR/$\Mh$ to the LBGs despite having $\sim 1$ dex lower SFR$/\Mh$. 
If these QGs have evolved from SBGs and QSOs like plotted here, the increase in $\MBH$ in QGs appears to be insignificant unless the $\fduty$ of SBGs and QSOs is less than $1\%$.

\vskip\baselineskip

We then compare $\NH$ among the four populations. The $\NH$ of our LBGs are predicted to be in the range $\log (N_\mathrm{H} / {\rm cm^{-2}}) = 23.25 - 23.96$ (see Section~\ref{sec: stacking_procedure}). The $N_\mathrm{H}$ of optically-selected QSOs (broad-line QSOs) are $\log (N_\mathrm{H} / {\rm cm^{-2}}) \lesssim 22$ \citep[e.g.,][]{Hasinger2008, Martocchia2017}, while the QGs have $\log (N_\mathrm{H} / {\rm cm^{-2}}) \sim 23.5$ 
\citep{Ito2022}. To estimate the $\NH$ of the SBGs, we perform an X-ray stacking analysis in a similar manner to the LBGs. No significant signal is detected at any redshift bin. By comparing the average
$3\sigma$ flux
upper limits 
thus obtained ($\log \langle F_\mathrm{X,stacked}^{3\sigma}\rangle$) with the average fluxes expected from super-deblended SED-fitting based AGN bolometric luminosities ($\log \langle F_\mathrm{X,expected}\rangle$),
we obtain lower limits of $\NH$ as shown in Table~\ref{tab:expecting_SBs_NH}. The SBGs have extremely high column densities of $\log (\NH / {\rm cm^{-2}}) \gtrsim 24.6-24.9$, being an order of magnitude higher than expected from \citet{Gilli2022}'s relation ($\log (N_\mathrm{H} / {\rm cm^{-2}}) =23.25-23.96$) and well into the Compton-thick regime. Thus, the $\NH$ of the four galaxy populations 
plotted here are different by at least $\sim 2.5$ dex 
although this large difference may be partly due to selection biases of the QSOs and SBGs.

The distribution of the four populations in Fig.~\ref{fig:sBHAR_vs_sSFR} and their $\NH$ values are qualitatively consistent with \citet{Hopkins2008}'s model in which major mergers of isolated SFGs end up with ellipticals after experiencing a dusty starburst phase with obscured AGN and a subsequent QSO phase.

\begin{table*}
        \begin{threeparttable}
	\caption{
 Expected $N_\mathrm{H}$ values of \citet{Jin2018}'s SBGs.
 }
	\label{tab:expecting_SBs_NH}
	\begin{tabular}{cccccccc} 
		\hline
            \multirow{2}{*}{$\langle z\rangle$} &
            $\langle $BHAR$_\mathrm{SED}\rangle$ $^{a}$ & 
            $\log \langle L_\mathrm{AGN,SED}\rangle$ $^{b}$ & 
            \multirow{2}{*}{$\log k_\mathrm{bol,SED}$ 
            $^{c}$} &
           $\log \langle L_\mathrm{X,SED}\rangle$ $^{d}$ & $\log \langle F_\mathrm{X,expected}\rangle$ $^{e}$ &
           $\log \langle F_\mathrm{X,stacked}^\mathrm{3\sigma}\rangle$ $^{f}$ &
           $\log N_\mathrm{H,expected}$ $^{g}$ \\
            & [$\Msun$ yr$^{-1}$] &
            [erg s$^{-1}$] &
               &
            [erg s$^{-1}$] &
            [erg s$^{-1}$ cm$^{-2}$] &
            [erg s$^{-1}$ cm$^{-2}$] &
            [cm$^{-2}$] 
            \\ \hline
            4 & 1.97 & 46.09  & 1.41  & 44.68  & $-$14.56  & $-$16.78 & 24.63 
            \\ 
            5 & 2.70 & 46.23  & 1.49  & 44.74  & $-$14.71  & $-$16.17 & 24.55  
            \\ 
            6.5 & 3.72 & 46.37  & 1.56  & 44.81  & $-$14.60  & $-$16.37 & 24.92 
            \\ \hline

	\end{tabular}

\textit{Notes}.

$^a$ Average BHAR derived from 
$\langle L_\mathrm{AGN,SED}\rangle$ $^b$.

$^b$ Average AGN luminosity 
calculated from \citet{Jin2018}'s catalog.

$^c$ X-ray to bolometric luminosity conversion factor calculated from 
\citet{Yang2018}'s Fig.8.

$^d$ Average AGN X-ray luminosity calculated as
$\langle L_\mathrm{AGN,SED}\rangle$ $^b$$/$$  k_\mathrm{bol,SED}$ $^c$.

$^e$ 
Expected average X-ray flux calculated from $\langle L_\mathrm{X,SED}\rangle$ $^d$ and $\langle z\rangle$.

$^f$ Average $3\sigma$ upper limit of $F_\mathrm{X}$ 
obtained by X-ray stacking analysis, after correction of the Galactic hydrogen
absorption.

$^g$ 
SBGs' $N_\mathrm{H}$ 
needed to achieve 
$\langle F_\mathrm{x,expected}\rangle$ $^e$.

\end{threeparttable}
\end{table*}

\subsection{Implications of Too High BHARs in Simulations}
\label{sec: simu-imply}
In Fig.~\ref{fig:simu_BHAR_vs_SFR}, we find that all four simulations predict too high BHARs.
The sub-grid models of BH accretion of these simulations use the Bondi-Hoyle-Lyttleton model (see Table 1 of \citet{Habouzit2021}):
\begin{align}
    \begin{split}
        \Dot{M}_\mathrm{Bondi} &= \frac{4\pi G^2 {M_\mathrm{BH}}^2 \rho}{c_\mathrm{s}^3},  
    \end{split}
    \label{eq:7}
\end{align}
or a modified version of it, 
where $G$ is the gravitational constant and
$\rho$ and $c_s$ are, respectively, the density and sound speed of the gas around the BH. SIMBA uses a torque-limited accretion model 
for cold gas as well. 

Because all simulations similarly overpredict the observed BHARs despite using different detailed formulations of accretion including the calculation of $\rho$ and $c_\mathrm{s}$, a possible cause of the overprediction would be that simulated SMBHs are too massive.  

\begin{figure*}
    \includegraphics[width=\textwidth]{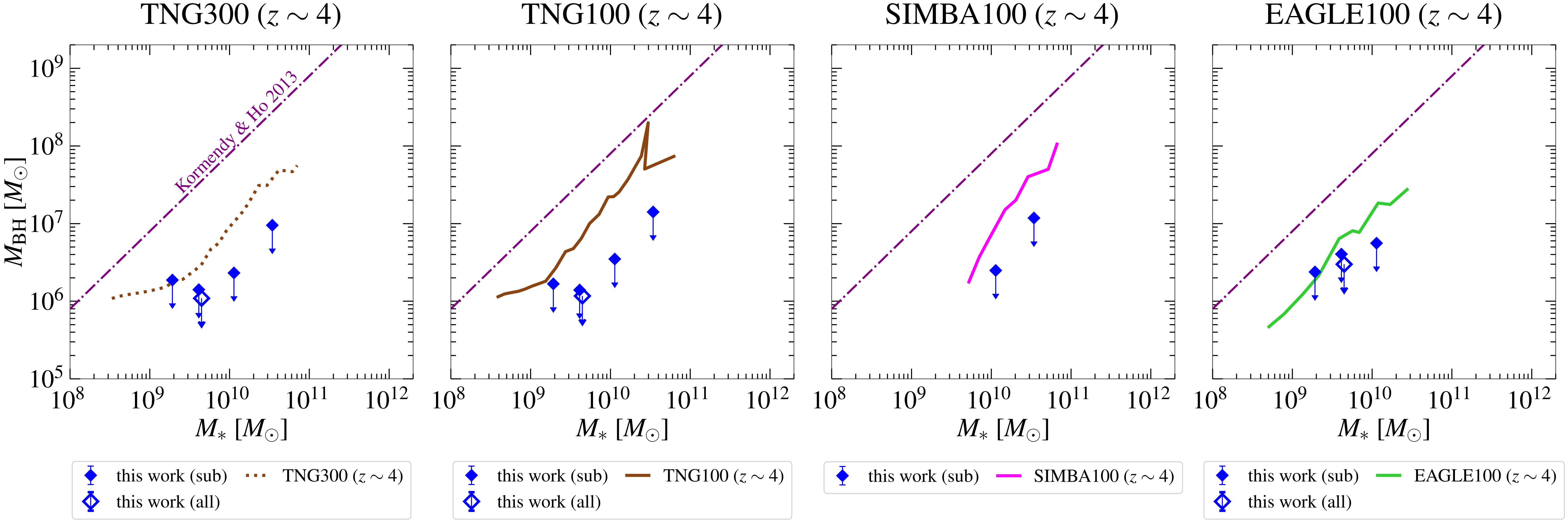}
    \caption{
    Comparison at $z\sim 4$ of simulated mean
    $\MBH-\Mstar$ relations (lines) with the $3\sigma$ upper limits of $\MBH$ 
    of our LBGs
    (diamonds with down arrows: open symbols are for total samples while filled symbols for magnitude divided samples) calculated on a simple assumption of $\langle$BHAR$\rangle^\mathrm{simul}/\langle$BHAR$\rangle^\mathrm{obs} = (\MBH^\mathrm{simul}/\MBH^\mathrm{true})^2$ for the four simulations (TNG300, TNG100, SIMBA100, and EAGLE100 from left).
    We adopt the same color choice as in Fig.~\ref{fig:simu_BHAR_vs_SFR} to represent each simulation. 
    Dot-dashed lines indicate 
    the local $\MBH$ -- $\Mbulge$ relation \citep{Kormendy2013}.
    In SIMBA100, the all, 25--26 mag, 
    and 26--27 mag
    samples are not plotted because SIMBA100 is incomplete at SFRs corresponding to these faint apparent magnitudes. Similarly, in EAGLE100, the 23--24 mag
    sample is not plotted because EAGLE100 does not contain such bright (i.e., high-SFR) galaxies.
    }
    \label{fig:MBH_vs_Mstar} 
\end{figure*}

On a simple assumption of BHAR$^\mathrm{simul}/$BHAR$^\mathrm{obs} = (\MBH^\mathrm{simul}/\MBH^\mathrm{true})^2$
(including SIMBA, in which the torque model for cold gas
dominates at high redshifts \citep{Habouzit2021}),
we calculate the average
$3\sigma$ upper limit of the true $M_\mathrm{BH}$ ($\MBHm$)
value at each of our $\BHAR$ data points.
Fig.~\ref{fig:MBH_vs_Mstar} shows the results for $z\sim4$ where the discrepancies of BHAR are largest. 
The true $\MBHm$ of our LBGs at $M_\star \gtrsim 10^{10} M_\odot$ are expected to be $> 3-5$ times lower than simulated.
Because the simulated $M_\mathrm{BH} - \Mstar$ relations are below the local relation, this result suggests that at $z\sim 4$, the SMBHs of SFGs are greatly undermassive compared with simultaneous evolution.
On the other hand, overmassive SMBHs have been found in high-redshift QSOs as mentioned in Section~\ref{sec:intro}. The distribution of $\Mbh$ at a fixed $\Mstar$ may be much wider than expected.
The discussion here is sketchy since simulated galaxies will not exactly follow $\mathrm{BHAR}\propto M_\mathrm{BH}^2$ even at fixed gas properties especially in SIMBA, and since decreasing $M_\mathrm{BH}$ will change gas properties accordingly.
In addition, we cannot rule out other causes. For example, the accretion rate decreases if the SMBH has a velocity, ${v}$, relative to the ambient gas, meaning that the denominator of eq.~\ref{eq:7} becomes $(
\mathit{c}_\mathrm{s}^2
+ v^2)^{3/2}$. 
Although ${v}$ is considered only in EAGLE \citep{Crain2015}, it might be non-negligible or underestimated.
A magnetic field in the gas may also reduce the accretion rate 
\citep[e.g.,][]{Kaaz2023}.
Investigating these causes is, however, beyond the scope of this paper.
In any case, the simulations appear to need some improvement to reproduce our low 
$\BHAR$.

\begin{figure*}
    \includegraphics[width=\textwidth]{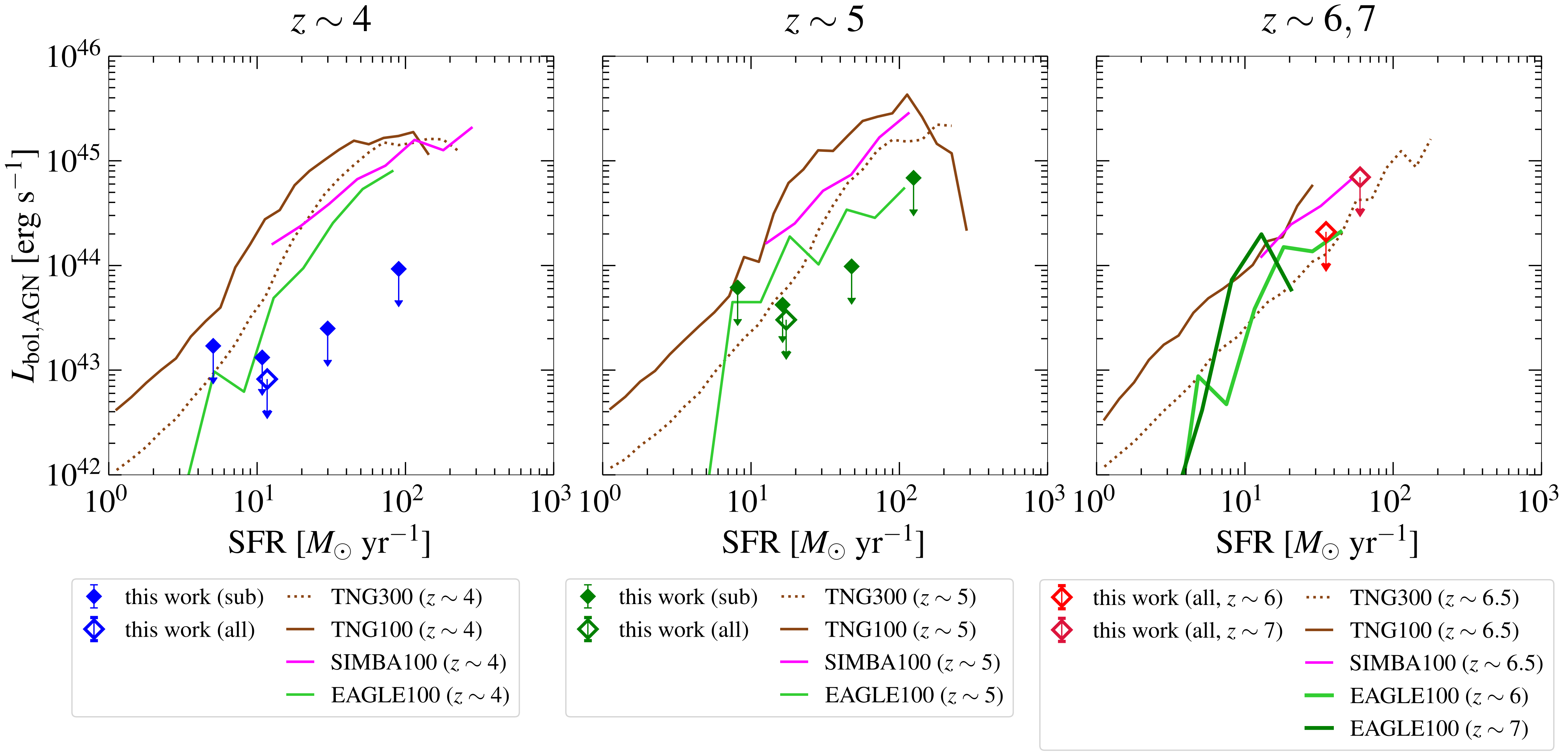}
    \caption{
    Mean $\LbolAGN$ versus SFR 
    predicted by the
    four simulations (lines: TNG300, TNG100, SIMBA100, and EAGLE100), compared with 
    the observed 
    $3\sigma$
    upper limits of 
    $\LbolAGN$ of
    LBGs (diamonds with down arrows: open symbols for total samples and filled ones for magnitude-divided samples).
    We calculate the $\LbolAGN$
    of simulated galaxies 
    from their BHAR taking into account the value of $\lamedd$ (see Section ~\ref{sec: simu-imply}).
    We adopt the same color choice as in Fig.~\ref{fig:simu_BHAR_vs_SFR} to represent each simulation.
    }
    \label{fig:Lbol_vs_SFR}
\end{figure*}

In the above comparison, we have adopted a constant radiative efficiency of 
$\epsilon=0.1$ or 0.2. 
However, $\epsilon$ may be lower for low activity (i.e., low Eddington ratio, 
$\lamedd=L_\mathrm{bol,AGN}
/L_\mathrm{edd}$, with $L_\mathrm{edd} = 1.26 \times 10^{38} (\MBH/M_\odot) \ {\rm erg\ s^{-1}}$)
systems. To examine the influence of such dependence on the comparison with simulations, we calculate $L_\mathrm{bol}$ for simulated galaxies considering $\lamedd$ dependence of $\epsilon$ and compare them with observed 
$\LbolAGNm$
values in Fig.~\ref{fig:Lbol_vs_SFR}. We have used the formula of $\epsilon(\lamedd)$ given by \citet{Churazov2005}, following \citet{Habouzit2021}:
\begin{enumerate}
    \item $\lamedd > 0.1$:
    \begin{align}
    \begin{split}
        \LbolAGN
        &= \frac{\epsilon}{1-\epsilon} \Dot{M}_\mathrm{BH} c^2,
    \end{split}
    \label{eq:lamedd>0.1}
\end{align}
    \item $\lamedd \le 0.1$:
    \begin{align}
    \begin{split}
        \LbolAGN
        &= 0.1L_\mathrm{edd}(10\lamedd)^2 = (10\lamedd)\epsilon \Dot{M}_\mathrm{BH} c^2.
    \end{split}
    \label{eq:lamedd<0.1}
\end{align}
\end{enumerate}

We find that the simulations again overshoot the data, although with slightly smaller discrepancies than seen in $\mathrm{BHAR}$. In fact, at $z\sim 4-7$, even inactive ($\mathrm{BHAR} \lesssim 1 M_\odot$ yr$^{-1}$) simulated galaxies like our LBGs have relatively high $\lamedd$.
Thus, the conclusions obtained from the comparison of BHARs with simulations are not sensitive to $\lamedd$ dependence of $\epsilon$.

\citet{Schirra2021} have 
found that all four simulations they have examined (Illustris, TNG100, EAGLE, and SIMBA) overestimate the $L_\mathrm{tot} (=L_\mathrm{X,AGN} + L_\mathrm{X,XRB})$ of SFGs 
at high redshift ($z\sim3, 4$) by comparing these simulations with \citet{Fornasini2018}'s SFG results in the $L_\mathrm{tot}$ -- $\Mstar$ plane.
\citet{Habouzit2022} have compared six cosmological simulations (Illustris, TNG100, TNG300, Horizon-AGN, EAGLE, and SIMBA) with observed AGN X-ray luminosity functions, finding that all but EAGLE produce too many AGNs with low X-ray luminosities of $L_\mathrm{X,2-10 keV} \sim 10^{43-44}$ erg $\mathrm{s^{-1}}$. 
These findings appear to be qualitatively in line with what we find.
These {\lq}overaccretion{\rq} problems may have the same origin.

\section{Summary}
\label{sec: summary} 
With an X-ray stacking analysis on a large ($N\simeq 12,000$) LBG sample, we have constrained the average BHAR of inactive SFGs at $4 \lesssim z \lesssim 7$ as functions of SFR and HAR. We have also compared the results with observations of QSOs, SBGs, and QGs at similar redshifts and with cosmological simulations of co-evolution.
Our main results and discussion are summarized as follows:
\begin{enumerate}
  \item At $z\sim 4$ and 5, the $3\sigma$ BHAR upper limits 
  of LBGs are 
  $\sim1.5$ dex and $\sim1$ dex 
  lower than expected from the local $\MBH/\Mstar$ ratio, respectively, implying that the SMBHs of SFGs at these redshifts are growing much more slowly than expected from simultaneous evolution. At $z\sim 6-7$, the upper limits are closer to the local mass ratio, possibly because of the small numbers of stacked sources.
  \item Similarly, at $z \sim 4$, LBGs' $\BHAR/\HAR$ is 
  $\sim 1-2$ dex
  lower than the $z=0$ $\MBH/\Mh$ ratio. The difference is smaller at higher redshifts, but the $\BHAR/\HAR$ is still $\sim 1$ dex lower at $z \sim 5$ and $\sim$ 0.5 dex lower at $z \sim 6-7$. 
  \item QSOs have $\sim$ 1 dex higher BHAR/SFR ratios than the $z=0$ $\MBH/\Mstar$ ratio, and hence $\sim 2$ dex higher ratios than the LBGs, implying that the SMBHs in QSOs are growing much faster than expected from simultaneous evolution. We find a similar trend in BHAR versus HAR.
   \item Effective (i.e., duty-cycle corrected) BHARs of $z\sim 4$ QSOs suggest that the $\MBH$ increase in the QSO phase dominates over that in the much longer, inactive ($\mathrm{BHAR} \lesssim 1\Msun$ yr$^{-1}$) period.
  \item QSOs and SBGs have $\sim 1$ dex higher SFR/$\Mh$ and $\gtrsim 2$ dex higher BHAR/$\Mh$ than LBGs, 
  suggesting that if the duty-cycle of QSOs' and SBGs' phases is $\gtrsim 10\%$, the 
  stellar mass
  acquired in these phases dominates over that by normal star formation.
  \item 
  SBGs have more enhanced star-forming activity than QSOs at a fixed BHAR$/\Mh$.
  The BHAR/SFR of SBGs is comparable to the local $\MBH/\Mstar$ ratio, meaning that their SMBH and stellar components are growing at paces roughly consistent with simultaneous evolution.
  \item QGs have comparable or higher BHAR$/\Mh$ to LBGs despite having $\sim 1$ dex lower SFR$/\Mh$. If these QGs have evolved from SBGs and QSOs, the increase in $\MBH$ in QGs appears to be insignificant unless the duty-cycle of SBGs and QSOs is $\lesssim 1\%$.
  \item The $\NH$ of our LBGs ($\log (\NH / {\rm cm^{-2}}) \sim 23.25 - 23.96$),  QSOs ($\log (\NH / {\rm cm^{-2}}) \sim 22$), SBGs ($\log (\NH / {\rm cm^{-2}}) \gtrsim 24.6-24.9$; compton-thick), QGs ($\log (\NH / {\rm cm^{-2}}) \sim 23.5$) differ at least $\sim 2.5$ dex.
  The locations of these galaxy populations in the BHAR/$\Mh$ versus SFR/$\Mh$ plane as well as their $\NH$ values are  
  qualitatively consistent with \citet{Hopkins2008}'s model in which major mergers of isolated SFGs end up with ellipticals after experiencing a dusty starburst phase with obscured AGN and a subsequent QSO phase.
  \item A comparison of our BHAR upper limits with four cosmological simulations (TNG300, TNG100, SIMBA100, and EAGLE100) finds that they overpredict the BHAR at a fixed SFR up to $\sim$1.5 dex. This result may imply that simulated SMBHs are too massive.
  Similar overpredictions are also found for $\Lbol$ calculated on an assumption that low-$\lamedd$ objects effectively have low $\epsilon$.
\end{enumerate}

The large differences in BHAR/SFR and BHAR/HAR between LBGs and QSOs, as well as LBGs' much lower BHARs than predictions of cosmological simulations, suggest diverse co-evolution paths at $z \gtrsim 4$, the era when the BHAD no longer follows the SFRD. We plan to utilize deep observational data of our LBGs by large telescopes such as the JWST 
(e.g., COSMOS-Web),
along with advanced simulations, to impose stronger constraints on co-evolution.

\section*{Acknowledgements}

We are grateful to the referee for the insightful comments
that have improved the paper.
We thank Francesca Civano and John Silverman for providing us with the Chandra Legacy Survey images.
We also thank Takuma Izumi and Yoshiki Toba for helpful discussions.
Our gratitude goes to Takamitsu Miyaji for his
generous support in using the \texttt{CSTACK} tool.
KS acknowledges support from JSPS KAKENHI Grant Number JP19K03924.
The scientific results
reported in this article are partially based on observations
made by the Chandra X-ray Observatory. This research has
used software provided by the Chandra X-ray Center (CXC) in
the PIMMS application package. 

\vspace{10pt}

\textit{Software}: CSTACK (v.4.4 \citealp{Miyaji2008}), PIMMS \citep{Mukai1993}, Astropy \citep{Robitaille2013, Price-Whelan2018}, Matplotlib \citep{Hunter2007},
numpy \citep{Harris2020}, pandas \citep{mckinney2010},
scienceplots \citep{Garrett2021}, h5py \citep{Collette2021}, yt \citep{Turk2011}, caesar (\href{https://caesar.readthedocs.io}{
https://caesar.readthedocs.io}).

\section*{Data Availability} 
The data from the TNG300 and the TNG100 simulations can be
found at \href{https://www.tng-project.org}{https://www.tng-project.org} and the data from the SIMBA simulation at \href{http://simba.roe.ac.uk/}{http://simba.roe.ac.uk/}. The data from the EAGLE100 simulation
can be obtained upon request to the EAGLE team at \href{http://icc.dur.ac.uk/Eagle/}{http://icc.dur.ac.uk/Eagle/}. The LBGs' catalog of \citet{Harikane2022} can be obtained from \href{http://cos.icrr.u-tokyo.ac.jp/rush.html}{http://cos.icrr.u-tokyo.ac.jp/rush.html}. The Chandra Legacy Survey catalog can be found at  \href{https://irsa.ipac.caltech.edu/data/COSMOS/tables/chandra/}{https://irsa.ipac.caltech.edu/data/COSMOS/tables/chandra/}. The SBGs' catalog of \citet{Jin2018} can be retrieved from \href{https://cdsarc.cds.unistra.fr/ftp/J/ApJ/864/56/}{https://cdsarc.cds.unistra.fr/ftp/J/ApJ/864/56/}.



\bibliographystyle{mnras}
\bibliography{matsui} 








\bsp	
\label{lastpage}
\end{document}